\documentclass[letterpaper,twocolumn,10pt]{article}
\usepackage{usenix}

\usepackage{amsmath}
\allowdisplaybreaks
\usepackage{booktabs}
\usepackage{caption}
\usepackage{graphicx}
\usepackage{amssymb}
\usepackage{diagbox}
\usepackage{xspace}
\usepackage{algpseudocodex}
\usepackage{bm}
\usepackage[utf8]{inputenc}
\usepackage{xcolor}
\usepackage{tikz}
\usepackage{pgfplots}
\usepackage{parnotes}
\usepackage{ragged2e}

\usetikzlibrary{shapes,
                arrows.meta,
                positioning,
                shadows,
                fadings,
                chains,
                patterns}
\usepackage{array}
\usepackage{amsthm}
\newtheorem{theorem}{Theorem}
\newtheorem{lemma}{Lemma}
\newtheorem{definition}{Definition}

\usepackage{multicol}
\usepackage{float}
\usepackage{fancybox}
\usepackage{color}  
\usepackage{algorithm}
\usepackage{algpseudocodex}

\usepackage[square, numbers, sort]{natbib}
\usepackage{tabularx}
\usepackage{nopageno}
\usepackage{enumitem}
\setlist[description]{leftmargin=1em}
\usepackage{balance}
\usepackage{threeparttable}
\usepackage{amsbsy}
\usepackage{listings}
\usepackage{tikzscale}
\newcounter{functionality}[section]

\newcounter{osimulator}[section]

\usepackage{mdframed} %

\usepackage[
    lambda, 
    advantage, 
    operators, 
    sets,
    adversary, 
    notions, 
    primitives, 
    asymptotics,
    landau,
    complexity]
{cryptocode}
\usepackage[capitalize]{cleveref}
\hypersetup{
  colorlinks=true,
  linkcolor={green!80!black},
  citecolor={red!60!black},
  urlcolor={cyan!70!black}
  }
\newtheoremstyle{noindentstyle} %
{3pt} %
{3pt} %
{\itshape} %
{0pt} %
{\bfseries} %
{.} %
{.5em} %
{} %
\usepgfplotslibrary{groupplots}
\pgfplotsset{every tick label/.append style={font=\small}}
\theoremstyle{noindentstyle}

\algrenewcommand\algorithmicreturn{\textbf{return}}

\makeatletter
\newenvironment{functionality}[1][htbp]
  {\refstepcounter{functionality} 
    \begin{algorithm}[#1]
    
    \floatname{algorithm}{Functionality}}
  {\end{algorithm}}
\crefname{functionality}{Func.}{Funcs.}
\Crefname{functionality}{Func.}{Funcs.}
\makeatother

\makeatletter
\newenvironment{osimulator}[1][htbp]
  {\refstepcounter{osimulator} 
    \begin{algorithm}[#1]
    
    \floatname{algorithm}{Oblivious Simulator}}
  {\end{algorithm}}
\crefname{osimulator}{Sim.}{Sims.}
\Crefname{osimulator}{Sim.}{Sims.}
\makeatother

\crefformat{section}{\S#2#1#3}
\crefmultiformat{section}{\S#2#1#3}{ and~\S#2#1#3}{, \S#2#1#3}{, and~\S#2#1#3}
\crefformat{subsection}{\S#2#1#3}
\crefmultiformat{subsection}{\S#2#1#3}{ and~\S#2#1#3}{, \S#2#1#3}{, and~\S#2#1#3}
\crefname{algocf}{Alg.}{Algs.} 
\Crefname{algocf}{Alg.}{Algs.}
\crefname{algorithm}{Alg.}{Algs.}
\Crefname{algorithm}{Alg.}{Algs.}
\crefname{ALC@unique}{Alg.}{Algs.}
\Crefname{ALC@unique}{Alg.}{Algs.}
\crefname{line}{line}{lines}
\Crefname{line}{Line}{Lines}
\crefname{definition}{Def.}{Defs.}
\Crefname{definition}{Def.}{Defs.}
\Crefname{tabular}{Tab.}{Tabs.}
\crefname{tabular}{Tab.}{Tabs.}
\Crefname{table}{Tab.}{Tabs.}
\crefname{table}{Tab.}{Tabs.}
\Crefname{figure}{Fig.}{Figs.}
\crefname{theorem}{Thm.}{Thms.}
\Crefname{theorem}{Thm.}{Thms.}
\crefname{lemma}{Lem.}{Lems.}
\Crefname{lemma}{Lem.}{Lems.}
\usepackage{subcaption}

\newcommand{\ie}{\textit{i}.\textit{e}.\xspace}
\newcommand{\eg}{\textit{e}.\textit{g}.\xspace}

\newcommand{\para}[1]{{\bf \noindent #1 }}

\newcommand{\system}{\ensuremath{\mathtt{H_2O_2RAM}}\xspace}
\newcommand{\codes}{\url{https://doi.org/10.5281/zenodo.14648338}\xspace}

\newcommand{\func}[1]{\ensuremath{\mathcal{F}_{\mathsf{#1}}}}

\pgfplotsset{compat=1.18}
\begin{document}

\title{\system: A High-Performance Hierarchical Doubly Oblivious RAM}

\author{
{\rm Leqian Zheng}\\
City University of Hong Kong
\and
{\rm Zheng Zhang}\\
ByteDance Inc.
\and
{\rm Wentao Dong}\\
City University of Hong Kong
\and
{\rm Yao Zhang}\\
ByteDance Inc.
\and
{\rm Ye Wu}\\
ByteDance Inc.
\and
{\rm Cong Wang}\\
City University of Hong Kong
} %

\maketitle

\begin{abstract}
The combination of Oblivious RAM (ORAM) with Trusted Execution Environments (TEE) has found numerous real-world applications due to their complementary nature.
TEEs alleviate the performance bottlenecks of ORAM, such as network bandwidth and roundtrip latency, and ORAM provides general-purpose protection for TEE applications against attacks exploiting memory access patterns. 
The defining property of this combination, which sets it apart from traditional ORAM designs, is its ability to ensure that memory accesses, both inside and outside of TEEs, are made oblivious, thus termed doubly oblivious RAM (O$_2$RAM). 
Efforts to develop O$_2$RAM with enhanced performance are ongoing. 

In this work, we propose \system, a high-performance doubly oblivious RAM construction.
The distinguishing feature of our approach, compared with the existing tree-based doubly oblivious designs, is its first adoption of the hierarchical framework that enjoys inherently better data locality and parallelization. 
While the latest hierarchical solution, FutORAMa, achieves concrete efficiency in the classic client-server model by leveraging a relaxed assumption of sublinear-sized client-side private memory, adapting it to our scenario poses challenges due to the conflict between this relaxed assumption and our doubly oblivious requirement.
To this end, we introduce several new efficient oblivious components to build a high-performance hierarchical O$_2$RAM (\system).
We implement our design and evaluate it on various scenarios.
The results indicate that \system reduces execution time by up to $\sim 10^3$ times and saves memory usage by a factor of $5\sim44$ compared with state-of-the-art solutions. 
\end{abstract}

\section{Introduction}\label{sec:introductino}
With the growing adoption of cloud computing, there have been rapidly arising concerns about the security and privacy of data outsourced to the cloud.
In this evolving landscape, trusted execution environments (TEEs) play an increasingly prevalent role due to their ability to provide enhanced security features (\eg, isolation, confidentiality, and integrity) without significant performance overhead, reliance on trusted third parties, or the need for non-colluding servers. 
Furthermore, virtual machine-based TEEs (\eg, Intel TDX~\cite{IntelTDX}, AMD SEV~\cite{AMDSEV}) require minimal adaption efforts to migrate existing applications into secure environments.
Consequently, many cloud providers~\cite{GoogleTEE, AzureTEE, AlibabaTEE} have incorporated TEEs as part of their infrastructure offerings. 

In brief, TEEs~\cite{Mechalas2016IntelSGX, IntelTDX, AMDSEV} provide high-level security by isolating the execution of selected code and data from the main operating system, thus shielding them even from system administrators.
Within the processor, a memory encryption engine transparently encrypts and decrypts confidential data as it moves to and from the main memory, using keys derived from a root key permanently embedded in the processor.
Furthermore, users can verify the integrity of the application or the system that operates within the TEEs via an attestation process~\cite{Attestation}, thus ensuring that computations are executed correctly and privately in the untrusted cloud.

\input{figs/orams}

However, many works~\cite{ChenCXZLL20,  BulckMWGKPSWYS18, LippKOSECG21, MurdockOGBGP20, LeeJFTP20, 0001SGKKP17, XuCP15, brasser2017software, Bulck2017TellingYS} have shown the shortcomings of TEEs that compromise user privacy. 
Some of them violating the intended security model (\eg, those from speculative execution~\cite{ChenCXZLL20,  BulckMWGKPSWYS18} and power analysis~\cite{LippKOSECG21}) are due to design or implementation flaws and will be patched by the corresponding manufacturers. 
In contrast, side-channel attacks~\cite{LeeJFTP20, 0001SGKKP17, XuCP15} related to memory access patterns generally remain beyond the security goal of mainstream TEEs, representing a persistent challenge.

This work focuses on concealing the memory access pattern that has been shown to be devastating in many applications~\cite{Islam2012AccessPD, Liu2014SearchPL, maas2013phantom, zhuang2004hide, Cash2015LeakageAbuseAA, Grubbs2016BreakingWA, xu2023leakage} in TEEs. 
Since the seminal work of \citet{goldreich1996software}, oblivious RAM (ORAM) has been recognized as a general and standard solution toward this goal. 
After decades of continuous development, ORAM has undergone significant advances with considerable efforts dedicated to performance optimization.
A promising line of work~\cite{mishra2018oblix, chamani2023graphos, EskandarianZ19, shih2017t, dauterman2021snoopy, zheng2017opaque, sasy2017zerotrace, shi2020path, tinoco2023enigmap} leverages TEEs to deploy a trusted client in the TEE to access the untrusted server with minimal latency and high bandwidth. 
This effectively mitigates one of the critical performance bottlenecks in the classic client-server model, \ie, network latency due to high bandwidth demands and round-trip complexity.
Meanwhile, the use of TEEs, in turn, brings a new consideration, \ie, ensuring that data accesses are oblivious both outside and within the TEE itself.
Thus, such solutions~\cite{mishra2018oblix, chamani2023graphos, sasy2017zerotrace, shi2020path, tinoco2023enigmap} are coined as \emph{doubly oblivious} RAM (DORAM).
To better distinguish it from distributed ORAM or differentially ORAM, we rename it as O$_2$RAM.

\para{ORAM roadmap.}Informally, ORAM conceals on which data blocks a client operates and whether these operations are reads or writes. 
This is achieved by accessing extra dummy blocks, shuffling the data periodically, and consistently writing the data back after each read. 
As discussed in~\cite{asharov2023futorama}, one prevalent category of tree-style ORAM designs, derived from Path ORAM~\cite{stefanov2018path}, excels in concrete efficiency, but falls short of achieving optimal theoretical complexity.
To get the value of a key $k$ in such designs as shown in~\cref{fig:orams}, one has to \emph{recursively} execute the following four steps until reaching the maximal possible height of a binary search tree, \eg, $\lceil 1.44\log_2 n\rceil$ for an AVL tree. 
{\normalsize \textcircled{\scriptsize 1}} starting from the root node, it obliviously compares $k$ with the current node value $v$ to determine whether the left/right node is the next one. 
{\normalsize \textcircled{\scriptsize 2}} to get $v$, it invokes Path ORAM's access protocol using the path identifier $id$ stored in the node. 
{\normalsize \textcircled{\scriptsize 3}} the access protocol retrieves a path by $id$ together with a randomly picked path. 
It linearly scans all the values retrieved and its stash to get the value $v$. 
Each node in Path ORAM, referred to as a bucket, contains $\sim 4$ data blocks, and the stash holds $\omega(\log n)$ blocks. 
{\normalsize \textcircled{\scriptsize 4}} afterwards, an eviction operation obliviously re-arranges the real data blocks over two retrieved paths and the stash as close to the leaves as possible w.r.t. the path invariance.

The other category derived from the foundational square-root ORAM~\cite{goldreich1987towards} draws more theoretical interest in the past.
It typically comprises $\log n$ levels, with each level increasing geometrically in size and structured as \emph{oblivious hash tables} (to be precise, oblivious w.r.t. non-recurrent lookups). 
To retrieve a value associated with $id$, the access protocol {\normalsize \textcircled{\scriptsize \romannumeral1}}, as shown in~\cref{fig:orams}, visits each nonempty level and re-writes the target data block back to the first one. 
A lookup in an oblivious hash table generally completes in $\bigO{\mathsf{poly} \log n}$ time.
{\normalsize \textcircled{\scriptsize \romannumeral2}} to maintain obliviousness, it extracts data from the first $i$ levels and rebuilds them into the $(i+1)$-th level after every $2^i$ accesses.
This rebuilding process, essentially relying on oblivious shuffling, dominates the computation overhead of hierarchical ORAM designs~\cite{PatelP0Y18PanORAMa, asharov2022optorama, asharov2023futorama}.
This line of work has recently achieved groundbreaking progress in achieving asymptotic optimality~\cite{asharov2022optorama, DittmerO20}. 
However, its practical efficiency is largely constrained by the huge constant hidden in the big-O notation.
\citet{asharov2023futorama} continue to optimize along this path, making it the first concretely efficient hierarchical scheme by allowing a sublinear yet reasonable-size client-side memory that does not expose access patterns. %

\para{Insight:}\emph{Under the premise of comparable complexity, hierarchical ORAM exhibits better concrete efficiency than tree-style ORAM due to its better data locality and parallelization.} 
As described above, data blocks are scattered along paths in tree-style ORAM, whereas in hierarchical ORAM, data blocks are continuously aligned at each level, facilitating better utilization of the cache and bandwidth of the chip. %
Meanwhile, hierarchical ORAM is more amenable to parallelization as the most time-consuming rebuilding process {\normalsize \textcircled{\scriptsize \romannumeral2}}, which mainly relies on oblivious shuffling, can be easily parallelized.
Although the eviction process in tree-style ORAM can also be partially parallelized, the volume of data involved is too small to justify the performance overhead caused by threading. 
Furthermore, while there are indeed some studies~\cite{williams2012privatefs, dauterman2021snoopy, ChakrabortiS19} exploring the parallelization and scaling of tree-style ORAMs, these efforts mainly focus on scaling \emph{a batch of} operations rather than parallelizing the internal computation for \emph{individual} operations.
Such a batch process technique is not ideal for many tasks, such as computing shortest paths, determining maximum flow, or data accesses with sequential oder. 

Given the achievable similar asymptotic complexity between hierarchical and tree-style ORAMs, we hence prefer the hierarchical one as our technical roadmap for further optimization.
In specific, our work builds upon FutORAMa~\cite{asharov2023futorama} that represents a leading advancement in hierarchical ORAM.
In essence, their approach assumes a reasonable yet non-constant, local memory that inherently does not expose access patterns (which does not apply to our scenario). 
In addition, they have implemented several elegant optimizations in the oblivious rebuilding and extraction processes of \emph{large} hash tables, approaching asymptotically optimal complexity.

\para{The challenge.}Developing an efficient doubly oblivious version of FutORAMa~\cite{asharov2023futorama} is non-trivial, as its high performance is heavily dependent on the use of sublinear-sized private memory that inherently conceals any access pattern. 
Without delving into excessive specifics, FutORAMa~\cite{asharov2023futorama} randomly distributes the data blocks at each level into bins (configured as hash tables), and then secretly selects a small subset of these blocks from each bin to a secondary and similarly configured structure.
The key point is that the client could leverage its private local memory to handle both \emph{small levels} and \emph{multiple bins} at large levels in a \emph{non-oblivious} (\ie, highly efficient) way.
A straightforward approach that replaces local memory with existing O$_2$RAM designs~\cite{mishra2018oblix, chamani2023graphos, sasy2017zerotrace} would incur substantial overhead.
Our empirical evaluation shows that even a simple look-up in such a na\"ive replacement (\ie, with a Path ORAM) incurs much more overhead than our final solution. 

Therefore, the primary challenge lies in constructing \emph{efficient} doubly oblivious bins, \ie, hash tables, that are of small to moderate size (concretely, less than $\sim 2^{18}$ data blocks or tens to hundreds of megabytes). 
This challenge arises for two main reasons:
1) each hash bin leveraged in large levels must be implemented efficiently and obliviously; 
2) while individual access times of small levels are significantly lower than those of large ones, optimizing their efficiency is also crucial as they are much more frequently accessed and rebuilt.

While several studies~\cite{kushilevitz2012security, chan2017obliviousCuckoo, goodrich2011cuckoo, PatelP0Y18PanORAMa, asharov2022optorama} have investigated efficient hash table designs for ORAM, most focus on optimizing large hash tables by decomposing them into several moderately sized ones, which are typically implemented as oblivious Cuckoo hash tables~\cite{chan2017obliviousCuckoo, goodrich2011cuckoo}.
These approaches are hence not applicable to our scenario.
As for the existing oblivious Cuckoo hashing schemes, they face two main issues: 1) a time-consuming oblivious build process; and 2) the necessity for a linear scan over a stash for a lookup to achieve negligible overflow or failure probability.
As acknowledged in~\cite{wang2014scoram_linear, doerner2017scaling_linear} and further confirmed in~\cite{noble2021explicit, mishra2018oblix}, the minimum size requirement of a stash significantly slows down the overall lookup of a Cuckoo hash table. 
This is because, for the oblivious access of a few dozen data blocks, a linear scan remains the most efficient method.
In concrete terms, existing Cuckoo hashing designs require a linear scan over \emph{hundreds or thousands} of data blocks for a \emph{single} operation, severely affecting their concrete performance.
In short, \emph{existing approaches are inadequate to construct doubly oblivious hash tables w.r.t. non-recurrent lookups that are concretely efficient, especially for those of small to moderate size}. 

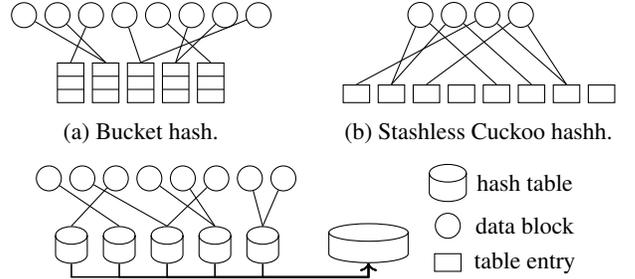
\begin{figure}[t]
    \centering
    
\begin{subfigure}[b]{0.47\linewidth}
    \centering
\begin{tikzpicture}[
    entry/.style={
        draw,
        circle,
    },
    bucket/.style={
        draw,
        rectangle,
        minimum height=1.5em,
        minimum width=1em,
        path picture={
            \pgfmathsetmacro{\numlines}{3}
            \newdimen\totalheight
            \newdimen\totalwidth
            \pgfextracty{\totalheight}{\pgfpointdiff{
                \pgfpointanchor{path picture bounding box}{south west}}
                {\pgfpointanchor{path picture bounding box}{north east}}}
            \pgfextractx{\totalwidth}{\pgfpointdiff{
                \pgfpointanchor{path picture bounding box}{south west}}
                {\pgfpointanchor{path picture bounding box}{north east}}}
            \pgfmathsetmacro{\tmp}{\totalheight/\numlines}
            \foreach \i in {1,...,\numlines} {
                    \draw (path picture bounding box.south west) ++(0,\i*\tmp *0.0351) -- ++(\totalwidth,0);
            };
        }
    },
]
\begin{scope}[start chain=going right,
              node distance=1mm]
\foreach \i in {1,...,8}
    \node[entry, on chain] (e1\i) {};
\end{scope}

\begin{scope}[start chain=going right,
              node distance=1mm,
              yshift=-0.9cm,
              xshift=6.2mm]
\foreach \i in {1,...,5}
    \node[bucket, on chain] (b1\i) {};
\end{scope}

\draw[-] (e11) -- (b12.north);
\draw[-] (e12) -- (b12.north);
\draw[-] (e13) -- (b11.north);
\draw[-] (e14) -- (b13.north);
\draw[-] (e15) -- (b15.north);
\draw[-] (e16) -- (b14.north);
\draw[-] (e17) -- (b14.north);
\draw[-] (e18) -- (b13.north);

\end{tikzpicture}
\caption{Bucket hash.}
\label{fig:ohashes:bucket}
\end{subfigure}
\hfill
\begin{subfigure}[b]{0.47\linewidth}
\centering
\begin{tikzpicture}[
    entry/.style={
        draw,
        circle,
    },
    bucket/.style={
        draw,
        rectangle,
        minimum height=0.5em,
        minimum width=1em,
    },
]
\begin{scope}[start chain=going right,
              node distance=1mm]
\foreach \i in {1,...,4}
    \node[entry, on chain] (e2\i) {};
\end{scope}

\begin{scope}[start chain=going right,
              node distance=1mm,
              xshift=-8.5mm,
              yshift=-1.05cm]
\foreach \i in {1,...,8}
    \node[bucket, on chain] (b2\i) {};
\end{scope}

\draw[-] (e21) -- (b22.north);
\draw[-] (e22) -- (b22.north);
\draw[-] (e23) -- (b21.north);
\draw[-] (e24) -- (b23.north);

\draw[-] (e21) -- (b25.north);
\draw[-] (e22) -- (b26.north);
\draw[-] (e23) -- (b27.north);
\draw[-] (e24) -- (b27.north);

\end{tikzpicture}
\caption{Stashless Cuckoo hash.}
\label{fig:ohashes:cuckoo}
\end{subfigure}

\vspace{0.5em}

\begin{subfigure}[b]{\linewidth}
\centering
\begin{tikzpicture}[
    entry/.style={
        draw,
        circle,
    },
    hash_table/.style={
        draw, 
        fill=white, 
        minimum height=1.2*0.4em,
        minimum width=1.2em, 
        cylinder,
        shape border rotate=90,
    },
    bucket/.style={
        draw,
        rectangle,
        minimum height=0.5em,
        minimum width=1em,
    },
]
\begin{scope}[start chain=going right,
              node distance=1mm]
\foreach \i in {1,...,8}
    \node[entry, on chain] (e\i) {};
\end{scope}

\begin{scope}[start chain=going right,
              node distance=2mm,
              yshift=-1cm,
              xshift=3mm,]
\foreach \i in {1,...,5}
    \node[hash_table, on chain] (b\i) {};
\end{scope}

\draw[-] (e1) -- (b2.north);
\draw[-] (e2) -- (b3.north);
\draw[-] (e3) -- (b1.north);
\draw[-] (e4) -- (b4.north);
\draw[-] (e5) -- (b4.north);
\draw[-] (e6) -- (b3.north);
\draw[-] (e7) -- (b5.north);
\draw[-] (e8) -- (b5.north);

\node[hash_table,
        minimum height=1em,
        minimum width=3em] (second_hash) at ($(b5.south) + (4em, .5em)$) {};

\foreach \i in {1,...,5}
    \draw[semithick, ->] (b\i.south) 
        -- ($(b\i.south) - (0, 0.5em)$)
        -- ($(b5.south) - (-4em, 0.5em)$)
        -- (second_hash.south);

\node[cylinder, 
      draw,
      fill=white,
      shape border rotate=90,
      minimum height=0.2em,
      minimum width=1.4em,] at ($(e8.south -| second_hash) + (3em, 0)$) (legend_t) {};
      
\node[entry] at ($(legend_t.south) - (0, 0.9em)$) (legend_e) {};
\node[bucket] at ($(legend_e.south) - (0, 0.9em)$) (legend_t_e) {};
      
\node at ($(legend_t.east) + (2.2em, 0.33em)$) (legend_t1) {\small hash table};

\node at ($(legend_e.west -| legend_t1)$) (legend_t2) {\small data block};

\node at ($(legend_t_e.west -| legend_t1)$) (legend_t2) {\small table entry};

\end{tikzpicture}
\caption{Two-tier hash, where a secret number of blocks from the main hash tables will be (obliviously) relocated to a secondary hash table.}
\label{fig:ohashes:two_tier}
\end{subfigure}

    \caption{Tailored hashing schemes used in this work.}
    \label{fig:ohashes}
\end{figure}

\para{Our approach.}To address the above issues, we tailor three oblivious hashing schemes, as shown in~\cref{fig:ohashes}, along with a na\"ive linear scan to handle varying scenarios. 

In specific, 
the bucket hash scheme, as shown in~\cref{fig:ohashes:bucket}, uses a pseudorandom function (PRF) to assign each block across buckets of uniform size. 
Each lookup requires a linear scan within the bucket identified by the PRF. 
The key aspect here is to derive the minimal bucket size to achieve a negligible bucket overflow probability caused by hash collisions. 
These bucket hash tables, detailed in~\cref{sec:ohashes:bucket}, are well-suited for managing small to moderately sized levels.

Our Cuckoo hash scheme, adapted from the elegant design~\cite{yeo2023cuckoo} as shown in~\cref{fig:ohashes:cuckoo}, removes the stash without sacrificing its negligible overflow probability. 
The key modification in the stashless design is simple yet powerful: use slightly more hash functions, each enjoying disjoint table entries.
However, there are still two technical issues to be addressed:
1) the work does not specify how to \emph{obliviously} build and access such hash tables. 
The existing designs for oblivious Cuckoo hash tables~\cite{goodrich2011cuckoo, chan2017obliviousCuckoo} are also not suitable for our scenario, as they are tailored for mechanisms with only two hash functions. 
To address this, we propose a new oblivious bipartite graph matching algorithm that may also be of independent interest in other domains.
We note that the complexity of this (relatively) expensive process remains independent of data block size, thus providing substantial benefits to our design in scenarios involving large blocks.
2) the asymptotic failure probabilities in~\cite{yeo2023cuckoo} are only tight for sufficiently large $n$, which deviates from our focus on moderate $n$. 
We hence derive a more concrete bound and employ numerical methods to compute more precise and relevant estimates, proving that $3\sim 6$ hash functions are sufficient to achieve a negligible failure probability.
These stashless Cuckoo hash tables, detailed in~\cref{sec:ohashes:cuckoo}, excel in scenarios where the number of lookups exceeds that of the data blocks they store.

The two-tier hash scheme, adapted from FutORAMa~\cite{asharov2023futorama} and illustrated in~\cref{fig:ohashes:two_tier}, closely resembles the bucket hash. 
Its performance benefits from the ability to safely place data blocks into major hash bins/tables in a \emph{non-oblivious} (\ie, efficient) way, given that 1) the input data are randomly shuffled, and 2) a small portion of each major bin is secretly and obliviously relocated to a secondary hash table. 
This approach effectively avoids the need for a relatively costly oblivious bin placement process.
Each lookup involves accessing a major bin identified by the PRF and the secondary hash table. 
Due to the minimum size requirement for the major bin to prevent both data overflow and underflow, this design is preferable for large levels. 
Further details are provided in~\cref{sec:ohashes:two_tier}.

In addition, we derive theoretical complexity analysis for the three oblivious hashing schemes mentioned above.
However, asymptotic analysis alone is insufficient for identifying the practically fastest scheme and parameters across various real-world scenarios, as factors such as parallelization, cache friendliness, and other physical considerations are hard to formalize and analyze as a whole to derive the optimal choice.
We hence develop a hash scheme planner, as discussed in~\cref{sec:planner}, to empirically select approximately optimal ones.\label{sec:intro:planner}

\para{Implementation and evaluation.}We implement, evaluate, and open source our O$_2$RAM design, along with some applications in doubly oblivious data structures and algorithms, \eg, O$_2$Map and single-source shortest paths. 
To the best of our knowledge, it is the first open-source O$_2$RAM design based on the hierarchical roadmap (the one in FutORAMa is implemented in Python and serves more as a simulation).
We compare our design with state-of-the-art tree-based O$_2$RAM designs, EnigMap~\cite{tinoco2023enigmap} for O$_2$RAM and O$_2$Map, and GraphOS~\cite{chamani2023graphos} for oblivious single-source shortest path computation.
The results show that \system outperforms \textsc{EnigMap}~\cite{tinoco2023enigmap} by factors up to $10^3\times$ in map operations. 
When integrated into specific applications, \system achieves speed-ups of $44\times$ in doubly oblivious single-source shortest path computations compared to GraphOS~\cite{chamani2023graphos}. 
In addition, \system also enjoys lower memory overhead, \eg, saving up to $44\times$ the memory space compared to \textsc{EnigMap}~\cite{tinoco2023enigmap}.

\para{Contributions.}We summarize our contributions as follows:
\begin{itemize}%
    \item We develop the first \emph{doubly oblivious} RAM based on the hierarchical roadmap, which not only shows concrete efficiency but also outperforms tree-based designs in practice. %
    \item Some of the building blocks, including oblivious stashless Cuckoo hashing, oblivious bipartite matching, and concrete parameter selection, are of independent interest.
    \item We provide an open-source implementation of \system on \codes along with empirical evaluations to show its concrete performance.  %
\end{itemize}

\section{Preliminaries}\label{sec:preliminaries}
\para{Balls into bins problem.}The process of throwing $m$ balls into $n$ bins in a uniformly random way perfectly simulates the data allocation of a bucket hash scheme, with only a negligible probability of deviation introduced by the PRF it uses. 
Our objective to find the minimum bucket size for a negligible overflow probability can be formalized as finding a threshold $k$ s.t. the probability of any bin that contains more than $k$ balls is negligible.
A general answer for this problem when $m=n$ is $k=\lceil e\cdot\log_2 n\rceil$.
However, this general answer is unsatisfactory for us to build an efficient bucket hash table.
We hence choose to state the following lemma to show a more precise overflow probability for finding a better bucket size:  
\begin{lemma}\label{lem:balls_into_bins}
If $n$ balls are thrown independently and uniformly into $m$ bins, the probability of a bin having at least $k$ balls is: 
$$
\Pr[\mathsf{overflow}] \leq \sum_{i=k}^{n} \binom{n}{i}\frac{(m-1)^{n-i}}{m^{n}}.
$$
\end{lemma}

\para{Bipartite graph and matching.}A bipartite graph is a graph  \(G = (L \cup R, E)\) where the vertex set \(V\) is divided into two disjoint sets \(L\) and \(R\), s.t. every edge connects a vertex in \(L\) and a vertex in \(R\) while no edge connects vertices within the same set. 
A \emph{matching} in a bipartite graph is a subset of edges \(M \subseteq E\) where no two edges share a common vertex. 
A matching is called left-perfect matching if the edges of $M$ cover all vertices in $L$. 

\para{Cuckoo hashing and its variants.}Cuckoo hashing~\cite{pagh2004cuckoo} in its simplest form has $2$ hash tables, $T_1$ and $T_2$, each with $m/2$ entries of unit capacity. 
The hash function $h_i$ for each table $T_i$ is chosen independently and is assumed to distribute inputs uniformly in $\{1, \dots, m/2\}$.
Given an item $x$, it is stored in either $T_1[h_1(x)]$ or $T_2[h_2(x)]$. 
To put it simply, we assign sequential indices from $1$ to $m$ to the entries of both hash tables.  
The outputs of hash functions are then adjusted with an appropriate offset.

As our focus is on scenarios where all input blocks are provided in advance rather than arriving online, constructing the Cuckoo hash table could be reduced to a bipartite matching problem.
Specifically, $n$ input data blocks and $m$ table entries form the disjoint vertex sets $L$ and $R$, respectively. 
Each hash function establishes an edge $(x, h_i(x))$ for each input data block $x$.  
It is clear that a matching directly corresponds to a table assignment and vice versa.

However, a left-perfect matching is not guaranteed to exist on such random graphs, indicating the input data blocks may overflow the hash table.
To address this, many variants have been proposed~\cite{yeo2023cuckoo, li2014algorithmic, noble2021explicit, FotakisPSS05, kirsch2010more}, including one of the most prevalent ones that place all overflowed data blocks in a stash. 
And it has been shown in~\cite{noble2021explicit, yeo2023cuckoo, FotakisPSS05} that the stash must be of $\Omega(\log n)$ size to prevent the overflowed blocks from exceeding its capacity. 
\citet{noble2021explicit} lists some concrete figures that for a single hash table of size $2^{16}$, the stash size has to be $\geq 14$ to obtain a $2^{-40}$ failure probability. 
As discussed in~\cref{sec:ohashes:cuckoo}, such stash sizes will bring notable performance drops.

Fortunately, \citet{yeo2023cuckoo} propose a novel variant that requires no stash but slightly more PRFs. 
In specific, it divides the entire table into $k$ sub-tables, each comprising $\lfloor m /k \rfloor$ entries. 
Every data block is then given $k$ candidate entries, one in each sub-table, selected by $k$ independent PRFs.
It is clear that the construction process is still equivalent to bipartite matching. 
The following lemma states its failure probability: 
\begin{lemma}[derived from~\cite{yeo2023cuckoo}]\label{thm:cuckoo_fail}
The failure probability of the above cuckoo hashing scheme with $n$ input data, $k$ independent PRFs, and $m$ table entries is upper bounded by: 
$$
\Pr[\mathsf{fail}]\!\leq\!\sum_{t=k+1}^{n}\binom{n}{t}\binom{m}{t-1}\left\lfloor\frac{m}{k}\right\rfloor^{-kt}\prod_{i=1}^{k}a_i^t,\text{ with} \sum_{i=1}^{k} a_i\!=\!t\!-\!1.
$$
\end{lemma}
The bound converges to $n^{-\Theta(k^2)}$ for {sufficiently} large $n$ and $k$. 
For practical usage, we employ numeric methods to compute appropriate $n$ and $k$, with detailed results provided in~\cref{sec:exp}. 
In brief, when $m=2n$, three to six hash functions are sufficient to reduce the failure probability to $2^{-64}$. 

\subsection{Threat Model}
We employ a threat model similar to those used in previous studies~\cite{mishra2018oblix, chamani2023graphos, sasy2017zerotrace} that integrate oblivious primitives with Trusted Execution Environments (TEEs). 
The key distinction between our model and the classic client-server model adopted in FutORAMa and other purely cryptographic designs lies in the fact that the client in the traditional model operates a fully trusted machine actively involved in the computation. 
In contrast, in our model, the client encrypts its data and uploads them to an untrusted server. 
The computation is then fully outsourced to the TEE, which physically resides on the untrusted server and may be exposed to various attacks.
Therefore, it is necessary to consider the presence of strong adversaries who can control the entire server’s software stack and the operating system, and even gain physical access to the server. 
While we do trust the secure processor and exclude cases in which an adversary could extract information from within the processor. 
We also assume that the TEE upholds its claimed security properties.
Although in reality, such an assumption may not always hold, just as in many other real-world applications, we can apply patches released by the manufacturer in a timely manner to maintain these standards.

In specific, we assume a remote attestation process~\cite{Attestation} that can help the client verify the identity and integrity of the TEE, and build a secure and authenticated communication channel. 
Data confidentiality is offered by a memory encryption engine within the processor that \emph{transparently} encrypts and decrypts private data as it travels to and from the chip cache, using keys derived from a root key burned during manufacturing.
Adversaries are also unable to tell whether two ciphertexts in and out the chip originate from the same plaintext.
We hence omit explicit descriptions of data encryption/decryption. 
In line with prior works~\cite{mishra2018oblix, chamani2023graphos, tinoco2023enigmap, sasy2017zerotrace, zheng2017opaque, EskandarianZ19}, we consider timing and power analysis attacks~\cite{LippKOSECG21}, rollback attacks~\cite{matetic2017rote}, and denial-of-service attacks as orthogonal issues. 
The interested readers may seek for works~\cite{Rane2015RaccoonCD, ShindeCNS16page1, Seo2017SGXShieldEA, shih2017t} for mitigations.

However, adversaries can observe \emph{which} memory data the TEE accesses, \ie, memory access patterns, using various methods and at different granularities. 
First, the untrusted host operation system can observe page-level access patterns via the page tables it maintains~\cite{AMDSEV, IntelTDX}.
A malicious system administrator can hence easily observe page-level access patterns. 
Second, shared resources, such as caches, can reveal which memory locations have been accessed. 
In addition, an attacker can mount an affordable hardware attack~\cite{LeeJFTP20} to obtain the exact data addresses accessed by the TEE by snooping on the physical memory bus.
It is hence essential to fully conceal the memory access patterns for general usage. 

\subsection{Oblivious Primitives}\label{sec:preliminaries:oblivious_primitives}
Due to space constraints, we defer some formal definitions to \cref{sec:append:oram}, but provide an overview of their concepts here.

\para{Obliviousness.}A RAM (or Turing machine) consists of a CPU containing a constant number of registers that are oblivious to adversaries, and a memory containing $n$ words indexed by $\{1, \dots, n\}$.
To execute a program that has some inputs and outputs, the CPU will interact with the memory in the form of reads and writes.
Adversaries can observe \emph{which} memory words the CPU accesses, named access pattern. 
We say that the program is oblivious if its access pattern is (statistically/computationally) independent of the content of input data. Namely, its access pattern can be simulated given only the input length.
We omit the formal definitions for the oblivious algorithms as they can be found in the referenced papers~\cite{asharov2023futorama, asharov2022optorama, PatelP0Y18PanORAMa}.
Meanwhile, we slightly abuse some terms by treating data blocks (items) the same as memory words of bit-length $w$. 
In implementation, we will consistently access the entire data block that may span several memory words.

\para{Oblivious operations.}We follow some standard assumptions~\cite{Sasy0G23Waks, mishra2018oblix, chamani2023graphos} that random bits can be generated obliviously. %
In our pseudocode, we leverage more readable descriptions like ``if $x>y$ then $a$ else $b$'', while they are implemented obliviously with some instruction-trace oblivious techniques~\cite{tinoco2023enigmap, liu2015oblivm, LiuHMHTS15}.

\para{Oblivious sorting/shuffling.}We adopt a classic yet concretely efficient sorting algorithm, bitonic sort, in our design. 
It achieves $\bigO{n\log^2 n}$ complexity for $n$ data blocks.
Oblivious shuffling is implemented via bitonic sorting with uniformly random keys.
A recent advancement, \textsc{WaksSort}~\cite{Sasy0G23Waks}, outperforms bitonic sort through an offline preprocessing phase. 
However, we choose not to adopt this design because the offline phase generates some randomness, resulting in considerably large space overhead as our work relies heavily on frequent invocation of oshuffle/osort.

\para{Oblivious compaction.}Given $n$ data blocks where some of the inputs are marked, it obliviously moves all the marked blocks to the front. 
There exist concretely efficient oblivious compaction algorithms~\cite{lin2019canosort, sasy2022fastocompact} taking $\bigO{n\log n}$ time. 
By further assuming that the input array is randomly shuffled with exactly half of the items marked, \citet{asharov2023futorama} achieve $\bigO{n}$ time with a negligible error probability $2n\exp(-Z/256)/Z$, where $Z$ denotes the local memory size. 
The original design compacts $Z$ items locally in $\bigO{Z}$ time. However, in our model, this process must be performed obliviously, introducing an additional $\bigO{\log Z}$ multiplicative overhead.
The total time hence becomes $\bigO{n\log Z}$.
When $n$ is small, the former is faster; otherwise, the latter is more efficient (a concrete threshold depends on the machine and block size). 
In our implementation, we always choose the better one.

\para{Oblivious intersperse.}Originated in~\cite{PatelP0Y18PanORAMa} and further optimized in~\cite{asharov2022optorama}, intersperse refers to the process of ``merging'' two randomly shuffled arrays into a single one that is also randomly shuffled. 
It is more efficient than na\"ively shuffling two arrays collectively and achieves the asymptotical optimality of $\bigO{n}$ time. 
In brief, it generates a random bit array, where each bit indicates whether the element at that location originates from the first or the second input array.
By reversing the process of \emph{obliviously compacting} the bit array, we can shuffle the items appropriately.
By slightly abusing the term, we also refer to it as ``oblivious shuffling''. 
Whether it is implemented as bitonic sort or intersperse, similar to the aforementioned ocompaction, depends on the specific context.

\para{Oblivious bin placement.}Given $n$ items, each tagged with a destination bin, %
the functional goal is to put items into their respective bins. 
Assuming $m$ bins of uniform size $k$, we adopt the method from~\cite{asharov2020bucket} with $\bigO{N\log^2 N}$ time, where $N:=n+m\cdot k$. 
In essence, it involves one round of osorting and one round of ocompaction over $N$ items.

\para{Oblivious hash table.}Given $n$ possibly dummy items in the form of $(k, v)$ pairs, an oblivious hash table supports the following three algorithms: 
\begin{itemize}[leftmargin=*,noitemsep]
\item \textsf{build} takes as inputs $n$ elements and creates a data structure; 
\item \textsf{lookup} receives a possibly dummy key $k$ and outputs the value $v$ corresponding to the key $k$ in the data structure or $\bot$ if $k$ is dummy or it is not found in the table; 
\item \textsf{extract} destructs the data structure and returns all elements that have never been looked up.
\end{itemize}

The security goal is to ensure that each of the three algorithms, both individually and in combination, remains oblivious.
Besides, we never call \textsf{lookup} with duplicate keys. 
In practice, we may represent distinct real keys as positive integers and dummies as negative integers, utilizing a dummy counter to avoid duplicate dummy lookups, under an input assumption of non-recurrent real keys.

\para{Hierarchical ORAM.}%
We opt to provide a detailed, end-to-end description of \system, adapted from FutORAMa~\cite{asharov2023futorama}, in~\cref{sec:append:oram}.
Instead, we provide a concise discussion of its process and components here, along with a simplified illustration of \system in~\cref{fig:orams}.
This discussion is for completeness and will also incorporate some minor modifications necessitated by our threat model.

Without loss of generality, we assume that the capacity $N$ of the ORAM (\ie, the maximum number of data blocks) is a two-power, otherwise, we can pad it to the next two-power. 
Let $L=\log N$, we create $L$ levels with each level $i$ instantiated as an oblivious hash table having a capacity of $2^i, \forall i\in\{1, \dots, L\}$ and initially marked as \emph{empty}. 
The bottom level, initiated with $N$ data blocks indexed from $1$ to $N$, is marked as non-empty.

In the access phase, given an operation $\mathtt{op}\in\{\mathtt{read}, \allowdisplaybreaks \mathtt{write}\}$, an address $\mathtt{addr}\in[N]$, and a data block $v\in\{0, 1\}^w$, we first initialize $\mathtt{res}:=\bot$, then for each nonempty level $i$: 
we lookup $\mathtt{addr}$ (or $\bot$) in its hash table if $\mathtt{res}=\bot$ (or $\mathtt{res}\neq\bot$); we then obliviously write $\mathtt{res}$ with the returned data if its key matches $\mathtt{addr}$. 
After the loop, we obliviously write $\mathtt{res}$ with $v$ if $\mathtt{op}=\mathtt{write}$. %
If the first level is full, we find the first empty level $i^*$ or set $i^*=L$ if all levels are full, then extract data from levels $1$ to $i^*-1$, 
and build $i^*$-th level with the extracted data as input.
Finally, we return $\mathtt{res}$ to the client.

\section{Oblivious Hash Schemes}\label{sec:ohashes}
As discussed in~\cref{sec:introductino,sec:preliminaries}, oblivious hash tables play a central and foundational role in hierarchical ORAM. 
In this section, we elaborate on the design of three tailored hash schemes as well as their parameter planners for better efficiency. 
Let $n$ denote the number of data blocks a hash table handles, and $t$ denote the total number of lookups over its lifetime.

\para{Concrete failure probability computation.}As described in~\cref{sec:preliminaries}, we opt to compute more precise overflow/error probabilities for better performance without compromising the system's security. 
Specifically, given that the computation primarily involves combinatorial and factorial numbers, we choose to compute their logarithms rather than their original values for better efficiency and numerical stability.
We will exponentiate the results back to their original values when necessary.
In addition, as the targeted overflow/error probabilities exceed native floating-point precision, we use high-precision mathematical libraries to ensure accurate results.

\subsection{Oblivious Bucket Hash}\label{sec:ohashes:bucket}
\para{Tight bucket size computation.}In oblivious bucket hash tables, we first have to determine the tight bucket size $\ell$ given $n$ and the number of buckets $m$. 
Equipped with~\cref{lem:balls_into_bins} and methods for calculating accurate probabilities, we can easily derive these sizes by binary search.

\para{Construction.}We then describe its construction as follows:

    \noindent$\bullet$\underline{$\mathsf{build}(A, m)$}: given $A$ containing $n$ input data blocks with unique keys and $m$ denoting the number of buckets, it begins by calculating the tight bucket size $\ell$ and initializing the table $T$ with $m$ empty buckets. 
    A PRF $\mathsf{PRF}$ with range $\{1, \dots, m\}$ and its key $\mathtt{sk}$ are sampled to compute the target bucket $\mathsf{PRF}_{\mathtt{sk}}(k_i)$ for $i$-th data block, after which the data in $A$ are obliviously placed into their target buckets (see~\cref{sec:preliminaries}). 
    
    \noindent\underline{$\mathsf{lookup}(k)$}: it performs a linear scan of the bucket identified by $\mathsf{PRF}_{\mathtt{sk}}(k)$ to obliviously select the matching data, which is subsequently marked as a dummy entry. 
    
    \noindent$\bullet$\underline{$\mathsf{extract}()$}: oblivious compaction is performed to retain exactly $n$ data entries, including all real ones.

\para{Time complexity.}The build process runs in the time complexity of $\bigO{(n+m\ell)\log^2(n+m\ell)}$ due to the oblivious bin placement operation. 
Assuming a hash table size of $m\ell=\Theta(n)$, it can be simplified as $\bigO{n\log^2 n}$.
Each lookup requires $\bigO{\ell}$ time due to the linear scan performed over the bucket.
The extraction mainly involves an ocompaction process, resulting in $\bigO{m\ell\cdot\log (m\ell)}$ time.
The overall time complexity for a bucket hash table handling $t$ lookups is hence $\bigO{(n+m\ell)\log^2(n+m\ell)+t\ell}$.

\para{Optimal bucket number.}So far, we have not yet specified how to determine the number of buckets $m$ for optimal running time. 
As discussed in~\cref{sec:intro:planner}, the above asymptotic analysis serves only as a reference and does not directly derive the optimal $m$. 
In fact, the actual running time, as shown in~\cref{fig:ohash_bucket_planner}, exhibits an approximately convex shape with significant fluctuations, where even slight changes in $m$ can lead to notable running-time variations.
Therefore, the highly time-consuming brute-force search is likely the only feasible method to find the optimal number of buckets.
However, if we relax the objective to an \emph{approximately} optimal solution, a variety of classic optimization methods can be employed.
In specific, we adopt the golden-section search~\cite{kiefer1953sequential} due to its simplicity.

\input{figs/obucket_hash}

\subsection{Oblivious Stashless Cuckoo Hash}\label{sec:ohashes:cuckoo}

We first claim that no solution practically surpasses na\"ive linear scans when dealing with small sets of data blocks. 
This has been empirically shown in several studies~\cite{wang2014scoram_linear, doerner2017scaling_linear, mishra2018oblix}. 
We supplement these findings by highlighting two critical factors that contribute to the observed deficiencies:
1) the oblivious build process typically involves several rounds of osorting over data sets that are two to three times larger than the input one;
and 2) the lower bounds on the number of entries that must be scanned during each lookup to achieve a negligible collision probability closely align with the size of the input data.
Concretely, for up to $256$ items, both bucket hash and Cuckoo hash with a stash~\cite{noble2021explicit} require scanning nearly $50$ entries per lookup. 
Similarly, a doubly oblivious map based on Path ORAM~\cite{mishra2018oblix, chamani2023graphos} requires a comparable number of scans, not to mention its higher building complexity.

The above finding places Cuckoo hash \emph{with} a stash in an awkward position in handling moderate-sized levels, as its stash typically holds hundreds of data blocks.
It is worth noting that combining stashes~\cite{goodrich2011cuckoo} across the entire ORAM does not effectively mitigate this issue, as the combined one is still relatively small. 
Fortunately, \citet{yeo2023cuckoo} proposed an elegant design that removes the need for a stash and proved its asymptotic optimality. 
However, making this design oblivious remains an unresolved challenge. 
To this end, we introduce an oblivious bipartite matching algorithm as follows. 
\subsubsection{Oblivious Bipartite Matching}
We remark that although our design resembles the existing oblivious Cuckoo hash build algorithm~\cite{chan2017obliviousCuckoo}, there are fundamental differences.
In their scheme, the two disjoint vertex sets comprise entries in the two hash sub-tables, with edges defined by $\left(h_1\left(x\right), h_2\left(x\right)\right)$ for each item $x$ where $h$ denotes PRFs. 
While in our design, the left vertex set consists of $n$ input items, the right vertex set consists of $2n$ table entries, with $k$ edges $(x, h_i(x)), i\in[k]$ for each item $x$.

\begin{figure}[b]
    \centering
    \begin{tikzpicture}[
    alter_edge/.style={
        solid, 
        draw,
        thick, 
        ->,
        -{Stealth[length=1.7mm]},
    },
    free_edge/.style={
        draw, 
        thick, 
        dashed,
        ->,
        -{Stealth[length=1.7mm]},
    },
    other_edge/.style={
        draw, 
        thick, 
        densely dotted,
        ->,
        -{Stealth[length=1.7mm]},
    },
]

\begin{scope}[start chain=going right,
              node distance=2.5mm]
\foreach \i in {1,...,5}
{
     \ifnum\i=1
        \node[circle, minimum size=4mm, draw, on chain, opacity=0.3] (L\i) {};
     \else 
        \ifnum\i=5
            \node[circle, minimum size=4mm, draw, on chain, cyan, thick] (L\i) {};
        \else 
            \node[circle, minimum size=4mm, draw, on chain] (L\i) {};
        \fi 
    \fi 
}
\end{scope}

\begin{scope}[start chain=going right,
              node distance=2.5mm, 
              yshift=-1.1cm]
\foreach \i in {1,...,5}
{
     \ifnum\i=1
        \node[circle, minimum size=4mm, draw, on chain, opacity=0.3] (R\i) {};
     \else 
        \node[circle, minimum size=4mm, draw, on chain] (R\i) {};
    \fi 
}
\end{scope}

\draw[alter_edge] (R5) -- (L4); 
\draw[alter_edge] (R4) -- (L3); 
\draw[alter_edge] (R3) -- (L2); 
\draw[alter_edge, opacity=0.3] (R1) -- (L1); 
\draw[other_edge, cyan] (L5)--(R5) node[midway, right] (l_l) {}; 
\draw[other_edge] (L3)--(R3);
\draw[other_edge] (L4)--(R4);
\draw[free_edge] (L2)--(R2);
\draw[free_edge, opacity=0.3] (L1)--(R2);

\begin{scope}[start chain=going right,
              node distance=2.5mm,
              xshift=0.5\linewidth]
\foreach \i in {1,...,5}
{
     \ifnum\i=1
        \node[circle, minimum size=4mm, draw, on chain, opacity=0.3] (Lb\i) {};
     \else 
        \node[circle, minimum size=4mm, draw, on chain] (Lb\i) {};
    \fi 
}
\end{scope}

\begin{scope}[start chain=going right,
              node distance=2.5mm, 
              yshift=-1.1cm,
              xshift=0.5\linewidth]
\foreach \i in {1,...,5}
{
     \ifnum\i=1
        \node[circle, minimum size=4mm, draw, on chain, opacity=0.3] (Rb\i) {};
     \else 
        \node[circle, minimum size=4mm, draw, on chain] (Rb\i) {};
    \fi 
}
\end{scope}

\draw[alter_edge, opacity=0.3] (Rb1) -- (Lb1) node[midway, left] (l_r) {}; 
\draw[free_edge, opacity=0.3] (Lb1)--(Rb2); 
\foreach \i in {2, ..., 5} 
    \draw[alter_edge] (Rb\i) -- (Lb\i); 
\foreach \i in {2, ..., 4} 
{
    \pgfmathtruncatemacro{\j}{\i+1}
    \draw[other_edge] (Lb\i) -- (Rb\j); 
}

\draw[draw, thick, -{Implies},double] (l_l) -- (l_r);

\node at ($(L1.west) - (1em, 0)$) {$L$};
\node at ($(R1.west) - (1em, 0)$) {$R$};

\begin{scope}[yshift=-1.5cm,] 
    \draw[alter_edge] (0,0) -- (1,0) node[pos=0, below, yshift=-0.7em, xshift=-0.4em, anchor=west]{\footnotesize alternating edge};
    \draw[free_edge] (2.5,0) -- (3.5, 0) node[pos=0, below, yshift=-0.7em, xshift=-0.4em, anchor=west]{\footnotesize free edge};
    \draw[other_edge] (5,0) -- (6,0) node[pos=0, below,yshift=-0.7em, xshift=-0.4em, anchor=west]{\footnotesize other edge};
\end{scope}

\end{tikzpicture}
    \caption{An example of bipartite matching: 
    the alternating/augmenting path is highlighted by the bold edges, all of which are reversed after inserting the blue vertex.}
    \label{fig:match}
\end{figure}
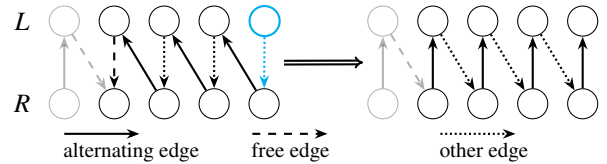

In fact, our algorithm is an oblivious version of the Hopcroft-Karp algorithm~\cite{HopcroftK73}. 
In brief, all edges within the matching are directed from right to left, while those outside the matching are directed from left to right.
We call a pair of edges that share the same vertex, with exactly one of them included in the matching, an alternating path.
With slightly abusing the terms, we also refer to the edges in the match as \emph{alternating} edges.
Adding a new item (\ie, a new vertex in $L$) is equivalent to \emph{recursively} finding an alternating path from the new vertex to any right vertex that is currently free.
For free vertices in $R$, we also name the edges incident to them \emph{free} edges.
We give a toy example of this process in~\cref{fig:match}.

\begin{algorithm}[htb]
    \caption{Oblivious bipartite matching $\mathsf{omatch}$}\label{alg:omatcher}
    \begin{algorithmic}[1] 
    \small
    \Require A bipartite graph $G=(L\cup R, E)$, a loop upper bound $\tau$;
    \Ensure A matching $M\subseteq E$.
    \For{$e=(u, v)\in E$}
    \State $e\gets \{(u,v),\ \mathtt{dir}:=r,\ \mathtt{st}:=\mathtt{unknown},\ \mathtt{ctr}:=0\}$\label{alg:omatch:tag}
    \EndFor 
    \For {$t\gets 0\dots, \tau$}
    \State obliviously sort $E$ by the rules in descending priority:\label{alg:omatch:osort1}
    \Statex\hspace{0.5em} 1. $e.u$ \Comment{{\small group edges incident to the same vertex in $L$}}
    \Statex\hspace{0.5em} 2. $e.dir=\ell$ \Comment{{\small  edges toward left first}}
    \Statex\hspace{0.5em} 3. $e.\mathtt{st}=\mathtt{free}$ \Comment{{\small  free edges first}}
    \Statex\hspace{0.5em} 4. $e.\mathtt{ctr}$ \Comment{{\small  edges with smaller counters first}} 
    \State initialize the matching result $M\gets \emptyset{}$, $e_0\gets \emptyset$
    \\
    {\color{gray}$\triangleright$\textit{\small all if-else clauses are implemented obliviously as introduced in~\cref{sec:preliminaries}}}
    \For{$e=(u, v)\in E$}\label{alg:omatch:loop1}
        \If{$e$ is the first edge in its group $e.u$}
        \If{$e_0.\mathtt{st}=\mathtt{alternat\_bk}$}
        \State $e_0.\mathtt{st}\gets\mathtt{alternat}$%
        \EndIf
        \State increment $e.\mathtt{ctr}\gets e.\mathtt{ctr}+1$ if $e.\mathtt{dir}=r$
        \State set $e.\mathtt{dir}\gets\ell, M[e_0.u]\gets e.v$, $e_0\gets e$ %
        \ElsIf{{\small $e.\mathtt{st}=\mathtt{free}$ and $e_0.\mathtt{st}=\mathtt{alternat}$}}
        \State set $e_0.\mathtt{dir}\gets r, e.\mathtt{dir}\gets\ell, M[e.u]\gets e.v$ %
        \State increment $e.\mathtt{ctr}\gets e.\mathtt{ctr}+1$
        \State update $e_0.\mathtt{st}\gets\mathtt{unknown}, e.\mathtt{st}\gets\mathtt{unknown}$
        \EndIf
    \EndFor
    \State obliviously sort $E$ by the rules in descending priority:\label{alg:omatch:osort2}
    \Statex\hspace{0.5em} 1. $e.v$ \Comment{{\small group edges incident to the same vertex in $R$}}
    \Statex\hspace{0.5em} 2. $e.dir=\ell$ \Comment{{\small  edges toward left first}}
    \Statex\hspace{0.5em} 3. $e.\mathtt{ctr}$ \Comment{{\small  edges with greater counters first}}
    \For{$e=(u, v)\in E$}\label{alg:omatch:loop2}
        \If{$e$ is the first edge in its group $e.v$}
        \State set $\mathtt{st} \gets \mathtt{free}$ if $e.\mathtt{dir}=r$ else $\mathtt{unkown}$
        \State set $e.\mathtt{st}\gets \mathtt{st}, e_0\gets e$
        \Else
        \State set $e.\mathtt{st}\gets \mathtt{st}$
        \If{{$e.\mathtt{dir}=\ell$}}
            $e_0.\mathtt{st}\gets \mathtt{altert\_bk}$, $e.\mathtt{dir}=r$
        \EndIf
        \EndIf
    \EndFor
    \EndFor 
    \State \Return $M$
    \end{algorithmic} 
\end{algorithm}

It is evident that the above recursive process will expose the structure of the bipartite matching, thereby exposing sensitive information of the input data.
Conceptually, our oblivious algorithm transforms this recursive process into a \emph{propagation}-based way.
Namely, a vertex in $L$ propagates its available \emph{free} and \emph{alternating} edges to the others via its neighbors in $R$. 
As shown in~\cref{alg:omatcher}~\cref{alg:omatch:tag} , we first tag each raw edge $e$ with three additional fields: 
1) $\mathtt{dir}\in\{\ell, r\}$ denotes its direction, 
2) $\mathtt{st}\in\{$\texttt{unknown}, \texttt{free}, \texttt{alternat}, \texttt{alternat\_bk}$\}$ denotes its state, 
and 3) $\mathtt{ctr}$ denotes the number of times it has been included in the match. 
Note that an edge directed to the left indicates that it is or tends to be included in the match.
We then iterate for a predefined and \emph{data-independent} times $\tau$. 
In each iteration, we obliviously sort the edges according to the rules shown in~\cref{alg:omatch:osort1}. 
It groups edges by their left vertices, and then edges toward the left first, followed by free edges toward the right. 
A counter is used to ensure that the edges previously tied in the back are moved to the front in each iteration.
The remaining ties are broken arbitrarily and consistently, \eg, using the right vertex. 
As shown in~\cref{alg:omatch:loop1}, we then try to include the first edge of each group in the match $M$. 
If the first edge tends to be alternated (\ie, reversed) and there do exist free edges, we will replace the old match by a new one (\ie, augmenting), and also update the state of both edges as \texttt{unknown}.

We then proceed to propagate edge states via vertices in $R$, which need to obliviously re-group edges based on their associated vertices in $R$ as shown in~\cref{alg:omatch:osort2}. 
Priority is given to edges that are oriented leftward and reversed more frequently within each group.
Then a linear scan shown in~\cref{alg:omatch:loop2} will mark the edges as free if $v\in R$ is not in the match, or preserve only one match for $v$ and correct the others. 
If there is more than one edge oriented leftward within a group, \ie, several left vertices are racing for $v$, we will mark the first winning one as \texttt{alternat\_bk}. 
Let $m$ be the size of the edge set and $d$ be the maximum possible length of augmenting paths, which are public to adversaries, we state the following theorem.
\begin{theorem}\label{thm:omatch}
    Assuming the existence of left-perfect matching and setting the number of iterations $\tau:=3d+1$, \cref{alg:omatcher} finds a left-perfect matching in $\bigO{\tau m\log^2 m}$ time.
\end{theorem}
\begin{proof}
    \Cref{alg:omatcher} runs in $\bigO{\tau m\log^2 m}$ time as it involves $\tau$ iterations, each dominated by two oblivious sortings of $\bigO{m \log^2 m}$ time. 
    Meanwhile, the obliviousness holds as two inner for-loops are implemented obliviously, and $\tau$ is independent of the input data. %
    The remaining challenge remains to prove its correctness.
    
    In the first iteration, all left vertices race for some right vertices and only a subset wins; 
    the winning edges are then marked as \texttt{alternat\_bk}, and some are labeled as \texttt{free}. 
    In subsequent iterations, all edges marked as \texttt{alternat\_bk} remain in the match and their states are changed to \texttt{alternat}; 
    edges already marked as \texttt{alternat} will ``augment'' if there are free edges in the same group, while vertices lost in the previous race attempt to find free edges or race for another round. 
    The strategy of state transitions from \texttt{alternat\_bk} to \texttt{alternat} effectively prevents vertices within the match from racing for free edges against vertices outside the match. 
    A vertex in the match will safely release a right vertex only if both of its new and old mates are not engaged in racing.
    And a vertex outside the match can possibly match the newly released vertex in the next iteration.
    Note that in the non-oblivious algorithm shown in~\cref{fig:match}, we recursively augment from the newly inserted vertex (i.e., racing vertices) to a free right vertex. 
    While in our algorithm, we progressively \emph{release} vertices or \emph{alternate} edges from the other end of the alternating path. 
    As it costs three iterations per alternation and alternations are performed concurrently, it takes at most $3d$ iterations to alternate all edges in the longest alternating path in addition to the first iteration. 
    Therefore, $3d+1$ iterations suffice to find a matching that is assumed to exist.
\end{proof}
\noindent For general bipartite graphs with $n:=|L|$, we have a simple collary \cref{alg:omatcher} finds a max-matching in $\bigO{nm\log^2 m}$ time as the maximum possible length of alternating paths is $n$.  

\subsubsection{Oblivious Stashless Cuckoo Hash Table}

Equipped with the oblivious bipartite matching algorithm, we are now prepared to present the construction of the oblivious stashless Cuckoo hash as shown in~\cref{alg:ocuckoo}.
To build the table from an input of $n$ data blocks and $k$ PRFs, we first establish the edges based on the keys of the data blocks, as shown in~\cref{alg:ocuckoo:graph}. 
We then run~\cref{alg:omatcher} to solve the allocation of input data, and subsequently place them obliviously (empty entries are filled with dummies).
For lookups, we scan $k$ table entries based on the PRF values for a real lookup, or $k$ randomly selected entries for a dummy lookup. 
Data extraction is performed obliviously by compacting real items to the front. %
\Cref{thm:ocuckoo} establishes its security and efficiency.

\begin{algorithm}[tbh]
    \caption{Oblivious stashless cuckoo hash \textsf{CHT}}\label{alg:ocuckoo}
    \begin{algorithmic}[1]
    \small
    \Statex\underline{$\mathsf{build}(A, k):$}
    \Comment{{\small $A$ contains $n$ input items $(k_i, v_i)$ with distinct keys}}
    \Statex \Comment{{\small $k$ denotes the number of hash functions to be used}}
    \State compute bucket size $b\gets \lfloor 2n/k\rfloor$
    \State sample a PRF with range $\{1, \dots, b\}$
    \State uniformly samples $k$ PRF keys $\mathtt{sk}_i$ at random
    \State initialize $L\gets\{1, \dots, n\}, R\gets\{1, \dots, 2n\}, E\gets\emptyset$
    \For{$i\gets 1, \dots, n$}\label{alg:ocuckoo:graph}
        \For{$j\gets 1, \dots, k$}
        \State compute $j$-th candidate entry $v\gets\mathsf{PRF}_{\mathtt{sk}_j}(A[i].k)+j\cdot b$ 
        \State establish an edge $e\gets (i, v)$ and $E\gets E\cup \{e\}$
        \EndFor 
    \EndFor 
    \State set $\tau\gets \max(3\log n+1, 30)$\label{alg:ocuckoo:tau}%
    \State find a matching $M\gets\mathsf{omatcher}((L\cup R, E),\tau)$
    \State parse $M$ as $(i, j)$, \ie, a bin placement scheme $P\gets (A[i], j)$
    \State obliviously place items in $A$ into $2n$ table entries $T$ w.r.t. $P$
    \State initialize a counter for dummy lookups $\mathtt{ctr}\gets 0$ 
    \end{algorithmic}
    
    \begin{algorithmic}[1]
    \Statex\underline{$\mathsf{lookup}(k'):$} \Comment{{\small assume that real keys are all possitive}}
    \State set $\mathtt{ret}\gets \bot$
    \State set $k'\gets -(\mathtt{ctr}+1)$ and increment $\mathtt{ctr}$ if $k'=\bot$
    \State compute entry indices $\mathtt{id}_i\gets \mathsf{PRF}_{\mathtt{sk}_i}(k'),$ $\forall i\in\{1, \dots, k\}$
    \State scan $T[\mathtt{id}_i], \forall i\in[k]$ and obliviously select the entry if $T[\mathtt{id}_i].k=k'$, and mark the matched one as dummy 
    \State \Return  $\mathtt{ret}$
    \end{algorithmic}
    
    \begin{algorithmic}[1]
    \Statex\underline{$\mathsf{extract}():$}
    \State obliviously compact the table $T$ and truncate to its half %
    \State obliviously shuffle $T$ uniformly at random
    \State \Return $T$
    \end{algorithmic}
\end{algorithm}

The key point in obliviously building a stashless Cuckoo hash table lies in bounding the iteration times $\tau$ (\cref{alg:ocuckoo:tau}) for \textsf{omatcher}, \ie, the maximum possible length of alternating paths. %
We adapt the proof technique in~\cite{FotakisPSS05}, establishing a bound of $\log n$ for somewhat large $n$ as shown in~\cref{lem:expansion,lem:tau_bound}. 
We also impose an explicit lower bound of $30$ for $\tau$ to handle scenarios involving small $n$.  

\begin{lemma}[derived from~\cite{FotakisPSS05}]\label{lem:expansion}
    Given the bipartite graph $G=(L\cup R, E)$ setup as in~\cref{alg:ocuckoo}, for $\mu\in[1, 2)$ and any $Y\subset R$ s.t. $|Y|=\mu n$, the following inequality holds with probability at least $1-e^{-\Omega(kn)}$:
    $$
    |\Gamma(Y)|\geq \frac{\mu}{2}n,
    $$
    where $\Gamma(Y)$ denotes neighbors of $Y$.
\end{lemma}
\begin{proof}
    Let $n$ denote the number of left vertices. 
    For $\mu\in[1, 2)$ and any $X\subset L$ with $|X|\geq (1-\mu/2)n$, we first prove that $|\Gamma(X)|\geq (2-\mu)n$ with all but negligible probability $e^{-\Omega(kn)}$ using same proof techniques in~\cite{yeo2023cuckoo}. 
    We then prove the lemma by contradiction. 
    Denote $X\gets \Gamma(Y)|$. 
    If $|X|<\frac{\mu}{2}n$, we will have $|X'| \geq (1-\mu/2)n$ with $X'\gets L\setminus X$. 
    Consequently, the number of neighbors of $X'$ must exceed $(2-\mu)n$, which implies that some vertices in $X'$ have neighbors in $Y$, contradicting to the definition of $X'$. 
\end{proof}

\begin{lemma}\label{lem:tau_bound}
    The maximum length of alternating paths in the random bipartite graph setup in~\cref{alg:ocuckoo} is bounded by $\log n$ with probability at least $1-e^{-\Omega(kn)}$. 
\end{lemma}
\begin{proof}
    We first assume the existence of a left-perfect matching $M$.
    Let $R_0$ be the set of free vertices in $R$, we have $|R_0|=n$ as $|R|=2n$ and $|M|=n$. 
    We expand $R_{\lambda}$ for $\lambda\geq 0$ as follows.  
    Let $L_{\lambda+1}\gets \Gamma(Y_{\lambda})$ be the neighbors of $Y_{\lambda}$ in $L$, and $R_{\lambda+1}\gets \Gamma(L_{\lambda+1}) \cup R_{0}$ be the neighbors of $L_{\lambda+1}$ together with free vertices, we claim that $|R_{\lambda}|\geq (2 -2^{-\lambda})n$, which can be derived from~\cref{lem:expansion} as $|R_{\lambda+1}|\geq n + |L_{\lambda+1}|\geq n+|R_{\lambda}|/2$ and $|R_0|=n$. 
    Namely, we have $|L_{\lambda+1}|\geq (1-2^{-\lambda-1})n$. 
    Taking $\lambda=\log n$ implies $|L_{\lambda +1}|\geq n-1$. 
    In brief, we can cover all vertices in $L$ in $\log n+1$ rounds of expansion, implying that the length of alternating paths is at most $\log n$. 
\end{proof}

\begin{theorem}\label{thm:ocuckoo}
Let $\beta$ be the size of the input data blocks.
\Cref{alg:ocuckoo} obliviously implements \func{HT} (formally defined in~\cref{func:oht}) for non-recurrent lookups with  $\bigO{nk\log^3 n+\beta n\log^2 n}$ building time, $\bigO{k}$ lookup time, and $\bigO{n\log^2 n}$ extraction time. 
In specific, the \textsf{build} process fails with a negligible probability $n^{-\Theta(k^2)}+e^{-\bigO{kn}}$. 
\end{theorem}

\begin{proof}
By \cref{thm:cuckoo_fail}, \textsf{build} fails with $n^{-\Theta(k^2)}$ probability due to the absence of a left-perfect matching. 
\Cref{lem:tau_bound} shows that we can find a left-perfect matching with all but $1-e^{-\bigO{kn}}$ probability. 
We hence obtained the claimed failure probability by combining them. 
Besides, as oblivious bipartite matching takes $\bigO{nk\log^3 n}$ time and the oblivious bin placement runs in $\bigO{\beta n\log^2 n}$ time, we then obtain the claimed building time complexity.

We build a simulator for the \textsf{build} process as follows. 
Given $n:=|A|$ and $k$ as inputs, the simulator samples $n$ key-value pairs with distinct keys as $A'$, and simulates \textsf{omatcher} and oblivious bin placement using their respective simulators.
The remaining steps are identical to the \textsf{build} process.
The parameter for the \textsf{omatcher} simulator, $\tau$, depends only on the number of input items $n$, which is public to adversaries. 
We can hence obliviously simulate the \textsf{build} process.

The \textsf{lookup} process is oblivious provided that the lookups are non-recurrent, and it runs in $\bigO{k}$ time as it linearly scans $k$ entries.
The running time of \textsf{extract} is dominated by an $\bigO{n\log^2 n}$ oshuffling. 
It is correct as at most half real data will be preserved by compaction.
We build a simulator for the \textsf{extract} process that obliviously simulates its functionality by substituting ocompaction and oshuffling with their respective simulators, while acknowledging that ``truncating-to-half'' operation is publicly known to adversaries. 
Note that substituting ocompaction with its simulator introduces a negligible failure probability~\cite{asharov2023futorama}.

\end{proof}

\para{Remarks on efficiency.}We observe that the edges introduced by the construction of Cuckoo hash tables can be safely divided into $k$ groups, each containing $n$ edges, as it is public to adversaries.
We can hence further optimize the oblivious matching algorithm by obliviously sorting each group of $n$ edges individually rather than collectively. 
This optimization slightly reduces the complexity of the \textsf{omatcher} algorithm from $\bigO{nk \log^2 nk}$ to $\bigO{k n\log^2 n}$.

Moreover, our proposed scheme not only enhances the lookup time compared to existing oblivious Cuckoo hash tables with a stash, but also reduces the building time by minimizing the rounds of required oblivious sorting in practice.
Another interesting property is that the running time of \textsf{omatcher}, one of the most time-consuming components in building the table, is independent of the size of the input data blocks. 
This independence offers a notable performance advantage when processing large data blocks.

\subsection{Oblivious Two-Tier Hash}\label{sec:ohashes:two_tier}
We follow the same framework as the ones in OptORAMa~\cite{asharov2022optorama} and FutORAMa~\cite{asharov2023futorama}. 
We omit its full formal description and focus only on our modifications as they are quite modular.
As shown in~\cref{fig:ohashes:two_tier} and discussed in~\cref{sec:introductino}, the key intuition behind is that we are safe to place data blocks into hash table buckets in a non-oblivious (\ie, efficient) way, as long as 1) they are randomly shuffled that can be efficiently accomplished by oblivious interspersion, 
and 2) a secret proportion, $\sim\epsilon$, of each major hash bucket/table is obliviously relocated to a secondary hash table, referred to as the overflow pile.

Our first modification is to implement both the major hash tables and the secondary hash table using our tailored hash tables, whereas the original designs employ oblivious Cuckoo hash tables with stashes. 
The specific hash scheme to be used in our two-tier hash tables will be determined, again, by our hash scheme planner as detailed in~\cref{sec:planner}.
Generally, the major hash tables use bucket hash, while the secondary hash table employs oblivious stashless Cuckoo hash. 
We emphasize that our Cuckoo hash scheme offers remarkable performance advantages for the overflow pile, as the number of accesses, $n$, to the overflow pile is several times greater than the amount of data, $\epsilon\cdot n$, it contains, and our Cuckoo hash scheme excels in lookup efficiency.
Concretely, it scans only several table entries, whereas a bucket hash needs to access dozens to hundreds of entries.

The second optimization is to adaptively select the ``overflow rate'' $\epsilon$ instead of using a fixed value. 
Although this does not yield better asymptotic complexity, it does enhance the actual runtime efficiency.
Note that we must keep $\exp(-\epsilon^2 Z/16)$ below a concretely negligible threshold~\cite{asharov2023futorama}, where $Z$ denotes the capacity of a major hash table. 
We then have $Z=C\cdot\epsilon^{-2}$ where $C$ is a constant (\eg, $1024$) to meet this condition.
As a meaningful $Z$ should be less than $n$, the possible values for $\epsilon$ are tightly constrained within $\left[\sqrt{C/n}, 1\right)$, particularly if we limit $\epsilon$ to powers of two.
Thus, a brute-force search is sufficient to find $\epsilon$ enjoying optimal running time. 

\para{Time complexity.}We assume that major hash tables adopt bucket hash and the overflow pile employs oblivious stashless Cuckoo hash. 
The building time for the two tiers are $\bigO{n\log^2 \left(\epsilon^{-2}C\right)}$ and $\bigO{\epsilon n\cdot\log^3(\epsilon n)}$, respectively.%
Note that two terms exhibit opposite monotonicities, \ie, as $\epsilon$ decreases, it takes more time to build the major hash tables while less time for the overflow pile, and vice versa.
The optimal value of $\epsilon$ depends on the specific block size and computing environments. 
For the sake of brevity, we assume $\epsilon=\log^{-3} n$, resulting in a total build time of $\bigO{n\log^2\log n}$, which is asymptotically equivalent to that of the bucket hash, but offers better concrete efficiency for large $n$. 
The \textsf{lookup} process, dominated by a major hash table's lookup, runs in $\bigO{\ell}$ time, where $\ell$ is the bucket size of major hash tables. 

We conclude this section with~\cref{tab:ohashes} that provides a clearer illustration of the hash tables used in \system. 
\begin{table}[htb]
    \centering
\begin{threeparttable}
    \caption{Summary of hash schemes used in \system.~\tnote{$\dagger$} }
    \label{tab:ohashes}
    \small 
    \begin{tabularx}{\linewidth}{m{0.25\linewidth}m{0.23\linewidth}m{0.14\linewidth}m{0.19\linewidth}}
        \specialrule{1.2pt}{1pt}{1pt}
        hash scheme & build \& extraction time~\tnote{$\ddagger$} & lookup time & suitable scenarios  \\
        \specialrule{0.7pt}{1pt}{1pt}
        linear scan & \multicolumn{2}{c}{$\bigO{n}$} & very small $n$  \\
        \specialrule{0.1pt}{1pt}{1pt}
        bucket hash~\cref{sec:ohashes:bucket}  & $\bigO{n\log^2 n}$ & $\bigO{\ell}$ & small to moderate $n$ \\
        \specialrule{0.1pt}{1pt}{1pt}
        stashless cuckoo hash~\cref{sec:ohashes:cuckoo}  & $\bigO{n\log^3 n}$ & $\bigO{1}$ & $n < t$\\
        \specialrule{0.1pt}{1pt}{1pt}
        two-tier hash~\cref{sec:ohashes:two_tier} & $\bigO{n\log^2\log n}$ & $\bigO{\ell}$ & large $n$ \\
        \specialrule{1.2pt}{1pt}{1pt}
    \end{tabularx}
    \begin{tablenotes}
    \item[$\dagger$] $n$: the size of a hash table, $\ell$: bucket size, $t$: the number of lookups performed during its lifetime.
    \item[$\ddagger$] Adopting an $\bigO{n\log n}$ osort algorithm~\cite{Sasy0G23Waks} saves a $\log n$ factor for this column.
    \end{tablenotes}
\end{threeparttable}
\end{table}

\section{\system}\label{sec:system}
We are now prepared to present \system, a high-performance hierarchical doubly oblivious RAM. 
As outlined in~\cref{sec:introductino}, \system builds on the hierarchical ORAM with its framework introduced in~\cref{sec:preliminaries:oblivious_primitives}. 
Due to space constraints, we defer the detailed description of \system and the proof of \cref{thm:h2o2ram}, which formalizes the security of \system, to \cref{sec:append:h2o2ram}.

\begin{theorem}\label{thm:h2o2ram}
    \system, formally described in~\cref{alg:h2o2ram}, obliviously implements ORAM functionality \func{ORAM} (defined in~\cref{func:oram}) with a negligible error probability. 
\end{theorem}

\para{Comparison with FutORAMa~\cite{asharov2023futorama}.}Hierarchical ORAM (HORAM), despite its long-standing history, has garnered more theoretical interest than practical adoption compared to the tree-based framework.
FutORAMa~\cite{asharov2023futorama} is the first to make the hierarchical framework practically viable. %
Though our work builds upon the HORAM framework, it differs from FutORAMa in the following aspects. %

The key difference lies in the threat model. 
FutORAMa operates in a classic client-server setting, where a client outsources its confidential data to an untrusted server and aims to hide data access patterns.
This approach demands intensive network communications, making its performance highly dependent on network conditions.
To make HORAM practical, FutORAMa introduces a key relaxation, enabling the client to manage a sublinear-sized private memory instead of being restricted to a constant-sized one. 
This allows the client to download a batch of data from the server, process it locally, and then send the processed data back, thereby reducing frequent network interactions.
In contrast, our work aims to address the limitation of TEEs in exposing memory access patterns. 
In TEEs, data access is not limited by the network conditions. 
However, most TEEs lack inherently oblivious memory, leaving us still constrained by the assumption of ``constant-sized client memory'' (\ie, registers within the CPU).
FutORAMa is hence incompatible with our setting. 

Besides, FutORAMa and our work target different optimization aspects, driven by the distinct threat models outlined above.
The optimizations in FutORAMa focus on efficiently utilizing sublinear-sized private memory while maintaining its reasonable size. 
Within this memory, \emph{non-oblivious} plain operations are performed.
Our work, on the other hand, concentrates on optimizing the small- and moderate-sized oblivious hash tables due to the lack of privileged memory in FutORAMa.
Note that while the hash tables we focus on are not large in size, their frequently accessed nature has a critical impact on the overall system performance, making it essential to optimize their efficiency.
To this end, we propose three hash schemes adapted and refined from existing works and develop a planner to determine the appropriate hash scheme for each level.
\system thus delivers substantial concrete performance improvements, despite having an asymptotic running time on par with that of the state-of-the-art approaches.

\para{Hash scheme planner.}As discussed in~\cref{sec:introductino,sec:ohashes}, it is impractical for us to calculate exact thresholds to identify the optimal hash scheme across various scenarios.
We hence develop a hash scheme planner to help achieve this goal.
First, it relies on the parameter selectors for each hash scheme, as outlined in~\cref{sec:ohashes}, to optimize the performance of these schemes.
It then selects the best one.
Of course, we could leverage the insights shown in~\cref{tab:ohashes} to minimize unnecessary comparisons. 
Namely, stashless Cuckoo hash will be considered a candidate only when used in the overflow pile of a two-tier hash table.
When a two-tier hash table outperforms a bucket hash table, subsequent levels will exclude the bucket hash scheme from consideration.

~\label{sec:planner}  %

\para{Amortized access time.}Denote the capacity of \system as $N$, w.l.o.g., we assume the bucket size $\ell=\bigO{\log N}$, \ie, a lookup runs in $\bigO{\log N}$ time. 
For $t$ accesses, we perform $\bigO{t\log N}$ lookups, and rebuild the $i$-th level $\lceil t/2^i\rceil$ times. 
Therefore, $t$ accesses require 
$
\bigO{t\log^2 N}+\sum_{i=1}^{\log N}\left\lceil \frac{t}{2^i}\right\rceil \bigO{2^i\log^2 \log 2^i}
$ time.
The amortized access time is hence $\tilde{O}\left(\log^2 N\right)$. 
We thus achieve a complexity comparable to tree-based O$_2$RAM designs~\cite{mishra2018oblix, tinoco2023enigmap, chamani2023graphos}, fulfilling the premise of our insight discussed in~\cref{sec:introductino}.
We also list an asymptotic comparison in~\cref{tab:orams}. 
\begin{table}[bht]
    \centering
    \caption{Asymptotic comparison of existing O$_2$RAM schemes, $N$ denotes the capacity.}\label{tab:orams}
    \small
    \begin{tabularx}{\linewidth}{m{0.5\linewidth}m{0.5\linewidth}}
    \toprule
        Scheme &  Access time  \\
        \midrule
        Oblix~\cite{mishra2018oblix} \& GraphOS~\cite{chamani2023graphos} & $\bigO{\log^3 N}$ \\
        \textsc{EnigMap}~\cite{tinoco2023enigmap} & $\tilde{O}(\log^2 N)$ \\
        \system & $\tilde{O}(\log^2 N)$ \\
    \bottomrule
    \end{tabularx}
\end{table}

\para{Extension for map support.}As noted in~\cref{sec:preliminaries}, a RAM of capacity $N$ is indexed by the logical address space $[N]$. 
However, in typical key-value map applications, keys are not confined to the range $[1, N]$. 
A na\"ive solution to support oblivious map is to implement a binary search tree with its backend data managed by ORAM with appropriate padding. %
While in our design, it is fortunate that all levels are implemented as (oblivious) hash tables,  making it irrelevant whether the keys fall within the range $[1,N]$ for correctness or security.
The only requirement is that the keys are hashed in a cryptographically secure and oblivious manner, and correct implementation of such hash functions is considered orthogonal to our work.

\para{Implementation.}We implement \system in C++ on \codes, and to the best of our knowledge, it is the first practical hierarchical-based O$_2$RAM implementation. 
Note that FutORAMa~\cite{asharov2023futorama} is implemented in Python and serves more as a simulation. 
Similar to FutORAMa, the initial level in \system does not begin with a capacity of two but a small linear scan level of capacity $256 \sim 1024$, depending on the actual block sizes.
The rationale is based on the fact that multiple linear scan levels are less efficient than a single combined one.

\begin{figure}[t]
    \centering
    \begin{subfigure}[b]{\linewidth}
        \centering
        \begin{tikzpicture}

    \definecolor{darkgray176}{RGB}{176,176,176}
    \definecolor{lightgray204}{RGB}{204,204,204}
    \definecolor{darkslateblue5759121}{RGB}{78,25,69}
    \definecolor{slateblue107110207}{RGB}{141,47,37}
    \definecolor{darkolivegreen9912157}{RGB}{203,148,117}
    \definecolor{darkkhaki181207107}{RGB}{144,146,145}

\begin{axis}[
legend cell align={left},
legend columns=-1, 
legend style={
  fill opacity=0,
  draw opacity=1,
  text opacity=1,
  at={(0.5, 1.28)},
  anchor=north,
  draw=black,
  fill=none,
  /tikz/every even column/.append style={column sep=3mm},
},
log basis x={2},
log basis y={2},
tick align=inside,
tick pos=left,
x grid style={darkgray176},
xlabel={\(\displaystyle \log_2(m)\)},
xmin=0.639128547547265, xmax=12096.1581667236,
xmode=log,
xtick style={color=black},
xtick={0.125,0.5,2,8,32,128,512,2048,8192,32768,131072},
xticklabels={
  \(\displaystyle {{-3}}\),
  \(\displaystyle {{-1}}\),
  \(\displaystyle {{1}}\),
  \(\displaystyle {{3}}\),
  \(\displaystyle {{5}}\),
  \(\displaystyle {{7}}\),
  \(\displaystyle {{9}}\),
  \(\displaystyle {{11}}\),
  \(\displaystyle {{13}}\),
  \(\displaystyle {{15}}\),
  \(\displaystyle {{17}}\)
},
y grid style={darkgray176},
ylabel={\(\displaystyle \log_2(\ell)\)},
ymin=17.9288447514547, ymax=10966.0160889088,
ymode=log,
ytick style={color=black},
ytick={4,16,64,256,1024,4096,16384,65536},
yticklabels={
  \(\displaystyle {{2}}\),
  \(\displaystyle {{4}}\),
  \(\displaystyle {{6}}\),
  \(\displaystyle {{8}}\),
  \(\displaystyle {{10}}\),
  \(\displaystyle {{12}}\),
  \(\displaystyle {{14}}\),
  \(\displaystyle {{16}}\)
},
width=\linewidth,
height=0.5\linewidth,
]
\addplot [very thick, darkslateblue5759121]
table {%
1 8192
2 4511
3 3128
4 2417
5 1982
6 1688
7 1475
8 1314
9 1187
10 1084
11 1000
12 929
13 868
14 816
15 771
16 731
17 695
18 663
19 634
20 609
21 585
22 563
23 544
24 525
25 509
26 493
27 479
28 465
29 452
30 441
31 429
32 419
33 409
34 400
35 391
36 382
37 375
38 367
39 360
40 353
41 346
42 340
43 334
44 328
45 323
46 318
47 313
48 308
49 303
50 299
51 294
52 290
53 286
54 282
55 278
56 275
57 271
58 268
59 265
60 261
61 258
62 255
63 252
64 249
65 247
66 244
67 241
68 239
69 236
70 234
71 232
72 229
73 227
74 225
75 223
76 221
77 219
78 217
79 215
80 213
81 211
82 209
83 208
84 206
85 204
86 203
87 201
88 199
89 198
90 196
91 195
92 193
93 192
94 191
95 189
96 188
97 187
98 185
99 184
100 183
101 182
102 180
103 179
104 178
105 177
106 176
107 175
108 174
109 172
110 171
111 170
112 169
113 168
114 167
115 166
117 165
118 164
119 163
120 162
121 161
122 160
123 159
124 158
126 157
127 156
128 155
129 154
131 153
132 152
133 151
135 150
136 149
137 148
139 147
140 146
142 145
143 144
145 143
147 142
148 141
150 140
152 139
153 138
155 137
157 136
159 135
161 134
162 133
164 132
166 131
168 130
170 129
173 128
175 127
177 126
179 125
182 124
184 123
186 122
189 121
192 120
194 119
197 118
200 117
202 116
205 115
208 114
211 113
215 112
218 111
221 110
224 109
228 108
232 107
235 106
239 105
243 104
247 103
251 102
255 101
260 100
264 99
269 98
274 97
279 96
284 95
289 94
294 93
300 92
306 91
312 90
318 89
325 88
331 87
338 86
346 85
353 84
361 83
369 82
377 81
386 80
395 79
404 78
414 77
425 76
435 75
446 74
458 73
470 72
483 71
497 70
511 69
525 68
541 67
557 66
574 65
592 64
611 63
631 62
653 61
675 60
699 59
724 58
751 57
779 56
810 55
842 54
877 53
914 52
954 51
996 50
1042 49
1092 48
1145 47
1203 46
1267 45
1335 44
1410 43
1492 42
1582 41
1681 40
1791 39
1913 38
2048 37
2200 36
2370 35
2562 34
2780 33
3028 32
3313 31
3643 30
4026 29
4476 28
5008 27
5643 26
6409 25
7345 24
};
\addlegendentry{$\delta=2^{-64}$}
\addplot [very thick, slateblue107110207, dashed]
table {%
1 8192
2 4689
3 3301
4 2578
5 2133
6 1829
7 1609
8 1441
9 1309
10 1201
11 1113
12 1038
13 974
14 919
15 870
16 828
17 790
18 756
19 725
20 697
21 672
22 649
23 628
24 608
25 590
26 573
27 557
28 543
29 529
30 516
31 504
32 492
33 482
34 471
35 462
36 453
37 444
38 436
39 428
40 420
41 413
42 406
43 400
44 393
45 387
46 381
47 376
48 370
49 365
50 360
51 355
52 351
53 346
54 342
55 338
56 334
57 330
58 326
59 322
60 319
61 315
62 312
63 309
64 305
65 302
66 299
67 296
68 293
69 291
70 288
71 285
72 283
73 280
74 278
75 275
76 273
77 271
78 269
79 266
80 264
81 262
82 260
83 258
84 256
85 254
86 253
87 251
88 249
89 247
90 245
91 244
92 242
93 240
94 239
95 237
96 236
97 234
98 233
99 231
100 230
101 229
102 227
103 226
104 224
105 223
106 222
107 221
108 219
109 218
110 217
111 216
112 215
113 213
114 212
115 211
116 210
117 209
118 208
119 207
120 206
121 205
122 204
123 203
124 202
125 201
126 200
127 199
128 198
129 197
130 196
132 195
133 194
134 193
135 192
136 191
137 190
139 189
140 188
141 187
143 186
144 185
145 184
147 183
148 182
149 181
151 180
152 179
154 178
155 177
157 176
158 175
160 174
162 173
163 172
165 171
167 170
168 169
170 168
172 167
174 166
176 165
178 164
180 163
182 162
184 161
186 160
188 159
190 158
192 157
194 156
197 155
199 154
201 153
204 152
206 151
209 150
211 149
214 148
217 147
220 146
222 145
225 144
228 143
231 142
234 141
238 140
241 139
244 138
247 137
251 136
254 135
258 134
262 133
266 132
270 131
274 130
278 129
282 128
286 127
291 126
295 125
300 124
305 123
310 122
315 121
320 120
325 119
331 118
337 117
343 116
349 115
355 114
361 113
368 112
375 111
382 110
389 109
397 108
404 107
412 106
421 105
429 104
438 103
447 102
457 101
466 100
477 99
487 98
498 97
510 96
521 95
534 94
546 93
560 92
573 91
588 90
603 89
618 88
635 87
652 86
669 85
688 84
707 83
728 82
749 81
771 80
795 79
819 78
845 77
872 76
901 75
931 74
963 73
997 72
1032 71
1070 70
1110 69
1152 68
1197 67
1245 66
1296 65
1350 64
1408 63
1470 62
1537 61
1608 60
1684 59
1767 58
1856 57
1952 56
2056 55
2169 54
2292 53
2426 52
2572 51
2732 50
2908 49
3102 48
3316 47
3554 46
3818 45
4113 44
4443 43
4815 42
5234 41
5711 40
6254 39
6877 38
7596 37
};
\addlegendentry{$\delta=2^{-128}$}
\addplot [very thick, darkolivegreen9912157, dashdotted]
table {%
1 8192
2 4939
3 3546
4 2808
5 2349
6 2034
7 1803
8 1626
9 1487
10 1373
11 1278
12 1198
13 1130
14 1070
15 1018
16 972
17 931
18 894
19 860
20 830
21 802
22 777
23 753
24 732
25 712
26 693
27 676
28 659
29 644
30 630
31 616
32 604
33 591
34 580
35 569
36 559
37 549
38 540
39 531
40 523
41 514
42 507
43 499
44 492
45 485
46 479
47 472
48 466
49 460
50 454
51 449
52 444
53 439
54 434
55 429
56 424
57 420
58 415
59 411
60 407
61 403
62 399
63 395
64 392
65 388
66 385
67 381
68 378
69 375
70 372
71 369
72 366
73 363
74 360
75 357
76 354
77 352
78 349
79 347
80 344
81 342
82 339
83 337
84 335
85 332
86 330
87 328
88 326
89 324
90 322
91 320
92 318
93 316
94 314
95 312
96 311
97 309
98 307
99 305
100 304
101 302
102 300
103 299
104 297
105 296
106 294
107 293
108 291
109 290
110 288
111 287
112 286
113 284
114 283
115 282
116 280
117 279
118 278
119 277
120 275
121 274
122 273
123 272
124 271
125 269
126 268
127 267
128 266
129 265
130 264
131 263
132 262
133 261
134 260
135 259
136 258
137 257
138 256
139 255
140 254
141 253
142 252
143 251
144 250
145 249
147 248
148 247
149 246
150 245
151 244
152 243
154 242
155 241
156 240
158 239
159 238
160 237
162 236
163 235
164 234
166 233
167 232
169 231
170 230
171 229
173 228
175 227
176 226
178 225
179 224
181 223
183 222
184 221
186 220
188 219
189 218
191 217
193 216
195 215
197 214
199 213
201 212
203 211
205 210
207 209
209 208
211 207
213 206
215 205
217 204
220 203
222 202
224 201
227 200
229 199
231 198
234 197
236 196
239 195
242 194
244 193
247 192
250 191
253 190
256 189
259 188
262 187
265 186
268 185
271 184
274 183
278 182
281 181
284 180
288 179
291 178
295 177
299 176
303 175
307 174
311 173
315 172
319 171
323 170
327 169
332 168
336 167
341 166
346 165
351 164
356 163
361 162
366 161
371 160
377 159
382 158
388 157
394 156
400 155
406 154
412 153
419 152
426 151
432 150
439 149
447 148
454 147
462 146
469 145
477 144
486 143
494 142
503 141
512 140
521 139
530 138
540 137
550 136
560 135
571 134
582 133
593 132
605 131
617 130
630 129
642 128
656 127
669 126
683 125
698 124
713 123
729 122
745 121
762 120
779 119
797 118
816 117
835 116
855 115
876 114
898 113
920 112
943 111
968 110
993 109
1019 108
1047 107
1075 106
1105 105
1136 104
1169 103
1203 102
1238 101
1275 100
1314 99
1355 98
1398 97
1442 96
1489 95
1539 94
1591 93
1645 92
1703 91
1763 90
1827 89
1895 88
1966 87
2041 86
2121 85
2206 84
2295 83
2391 82
2492 81
2600 80
2715 79
2837 78
2968 77
3108 76
3258 75
3419 74
3592 73
3778 72
3979 71
4195 70
4429 69
4682 68
4957 67
5255 66
5580 65
5935 64
6323 63
6749 62
7216 61
7731 60
};
\addlegendentry{$\delta=2^{-256}$}
\end{axis}

\end{tikzpicture}
        \caption{Bucket size $\ell$ versus the number of buckets $m$ required to achieve an overflow probability $\delta$ in bucket hash with $8192$ data blocks.}\label{fig:fail_prob:bucket}
    \end{subfigure}
    
    \begin{subfigure}[b]{\linewidth}
        \centering
        \begin{tikzpicture}

\definecolor{crimson2143940}{RGB}{214,39,40}
\definecolor{forestgreen4416044}{RGB}{44,160,44}
    \definecolor{darkgray176}{RGB}{176,176,176}
    \definecolor{lightgray204}{RGB}{204,204,204}
    \definecolor{darkslateblue5759121}{RGB}{78,25,69}
    \definecolor{slateblue107110207}{RGB}{141,47,37}
    \definecolor{darkolivegreen9912157}{RGB}{203,148,117}
    \definecolor{darkkhaki181207107}{RGB}{144,146,145}
\begin{axis}[
legend cell align={left},
legend columns=-1, 
legend style={
  fill opacity=0,
  draw opacity=1,
  text opacity=1,
  at={(0.5, 1.28)},
  anchor=north,
  draw=black,
  fill=none,
  /tikz/every even column/.append style={column sep=3mm},
},
log basis x={2},
log basis y={2},
tick align=inside,
tick pos=left,
x grid style={darkgray176},
xlabel={$\log_2 n$},
xmin=18.3791736799526, xmax=3651353.70983519,
xmode=log,
xtick style={color=black},
xtick={16,256,4096,65536,1048576,16777216},
xticklabels={4,8,12,16,20,24},
y grid style={darkgray176},
ylabel={$\log_2 \delta$},
ymin=9.40323380520589e-164, ymax=109830.537933332,
ymode=log,
ytick style={color=black},
yticklabel style={},
ytick={5.4210108624275221700372640043497e-20,2.9387358770557187699218413430556e-39,8.6361685550944446253863518628004e-78},
yticklabels={-64,-128,-256},
width=\linewidth,
height=0.45\linewidth
]
\addplot [very thick, darkslateblue5759121]
table {%
32 0.00251938043687018
64 3.52998298241345e-05
128 9.18863038926687e-07
256 2.902317594221e-08
512 8.82928083741408e-10
1024 2.77191641528542e-11
2048 8.60657174494347e-13
4096 2.69272922296726e-14
8192 8.4012602931434e-16
16384 2.6261683117039e-17
32768 8.20347817090845e-19
65536 2.56377588984752e-20
131072 8.01099465603067e-22
262144 2.50348195413705e-23
524288 7.82318458779341e-25
1048576 2.4447564189265e-26
2097152 7.63981594860815e-28
};
\addlegendentry{$k=3$}
\addplot [very thick, slateblue107110207, dashed]
table {%
32 1.17723479845952e-07
64 7.01198638236677e-11
128 3.79101208842431e-14
256 1.94756693030711e-17
512 9.75414896573318e-21
1024 4.82361341587145e-24
2048 2.37027715347428e-27
4096 1.16104086714659e-30
8192 5.67814828030615e-34
16384 2.77473407802959e-37
32768 1.35538823602508e-40
65536 6.61941954232754e-44
131072 3.23245903319009e-47
262144 1.57842741428923e-50
524288 7.70735618162978e-54
1048576 3.76340418452899e-57
2097152 1.83761113020364e-60
};
\addlegendentry{$k=4$}
\addplot [very thick, darkolivegreen9912157, dash pattern=on 1pt off 3pt on 3pt off 3pt]
table {%
32 3.24117945207011e-11
64 2.53345239829597e-17
128 3.13572370050617e-23
256 6.47582414410746e-29
512 1.28481474400368e-34
1024 2.3221829630689e-40
2048 4.31180612779231e-46
4096 8.26441777325496e-52
8192 1.58016787843312e-57
16384 3.00382915926229e-63
32768 5.71974077387992e-69
65536 1.09128698124896e-74
131072 2.08178237294587e-80
262144 3.96985176812128e-86
524288 7.57109702424831e-92
1048576 1.44409970851686e-97
2097152 2.75442785647689e-103
};
\addlegendentry{$k=5$}
\addplot [very thick, darkkhaki181207107, dotted]
table {%
32 1.14172466534758e-15
64 4.45550784281729e-25
128 1.04720471160008e-33
256 1.32953272052729e-42
512 2.61977431233635e-51
1024 4.43584662662574e-60
2048 8.37850240809665e-69
4096 1.52389331396864e-77
8192 2.84837132799115e-86
16384 5.27402297420291e-95
32768 9.83218256041896e-104
65536 1.82866406020624e-112
131072 3.40689298029372e-121
262144 6.34347287740327e-130
524288 1.18162813144765e-138
1048576 2.20074925630822e-147
2097152 4.09927064616588e-156
};
\addlegendentry{$k=6$}
\addplot [thick, black, opacity=0.4, dotted, forget plot]
table {%
18.3791736799526 5.42101086242795e-20
3651353.70983518 5.42101086242795e-20
};
\addplot [thick, black, opacity=0.4, dotted, forget plot]
table {%
18.3791736799526 2.93873587705572e-39
3651353.70983518 2.93873587705572e-39
};
\addplot [thick, black, opacity=0.4, dotted, forget plot]
table {%
18.3791736799526 8.63616855509478e-78
3651353.70983518 8.63616855509478e-78
};
\end{axis}

\end{tikzpicture}
        \caption{Concrete failure probabilities of stashless Cuckoo hash, $\delta$, for various hash table sizes $n$ and numbers of hash functions, $k$.}\label{fig:fail_prob:cuckoo}
    \end{subfigure}
    \caption{More precise overflow/failure probabilities.}\label{fig:fail_prob}
\end{figure}
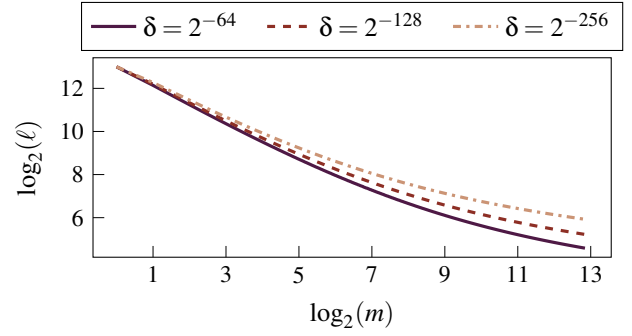
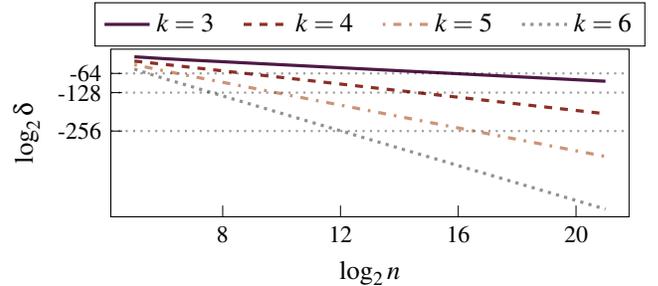

\section{Experimental Evaluation}\label{sec:exp}

\para{Experimental setup.}We implement \system with C++20, and measure the execution time using Google Benchmark~\cite{googleBenchmark}. 
The experimental setup involves a physical server powered by a 96-core Intel Xeon(R) Platinum 8457C processor and 2TB of RAM. 
Each core operates at a frequency of 2.6 GHz. 
This server utilizes Debian GNU/Linux 10 as its operating system, with the kernel upgraded to version 5.15.120+. 
Using software packages provided by Intel Trusted Domain Extension (TDX)~\cite{IntelTDX}, a confidential virtual machine is configured with 64 cores, 512 GB RAM, and ubuntu-20.04, which has the same version kernel as the host. 
We use both OpenMP~\cite{openmp} and Intel oneTBB~\cite{tbb} libraries to parallelize \system. 
All TEE experiments discussed throughout our work are conducted within this trusted VM.
To better measure the amortized access time of \system, we access both O$_2$RAM and O$_2$MAP for $n$ times and report the average access time. 

\para{Compared approaches.}We compared O$_2$MAP access times with \textsc{EnigMap}~\cite{mishra2018oblix}, and doubly oblivious single-source shortest path computation with GraphOS~\cite{chamani2023graphos}. 
Both baseline approaches stand for state-of-the-art tree-based doubly oblivious solutions.
Since TDX provides ample protected memory, we removed the manual encryption and decryption components from the \textsc{EnigMap} implementation~\cite{enigmap_code}, \ie, relying solely on TDX's encryption mechanisms. 
As a result, our experiments are expected to perform faster than those on SGX (note that our CPU is more powerful).
However, our reproduced results for \textsc{EnigMap} are approximately ten times slower than those reported in~\cite{tinoco2023enigmap}. 
Nonetheless, we confirm that their results still significantly surpass other state-of-the-art methods (\eg, \textsc{Omix}++~\cite{chamani2023graphos}) and do not undermine their conclusions presented in their abstract.

\subsection{Experiment Stages}

\para{Concrete overflow/failure probability.}As discussed in~\cref{sec:ohashes}, we leverage more precise formulas together with numeric methods and high-precision math libraries to calculate concrete error probabilities, allowing a selection of more optimal parameters.
\Cref{fig:fail_prob:bucket} shows tight bucket sizes to ensure a (concretely) negligible overflow probability. 
We note that our method yields notably smaller bucket sizes compared to general yet loose bounds, such as $e\log_2 n$~\cite{balls_into_bins}, and $267$~\cite{asharov2023futorama}. 
As shown in~\cref{fig:fail_prob:bucket}, adding just one more hash function to the basic Cuckoo hash scheme is sufficient to achieve a relatively low failure probability $2^{-64}$ for hash tables larger than $2^{15}$.
For smaller hash tables and stronger security (\ie, lower failure probabilities), $4$ to $6$ hash functions suffice.

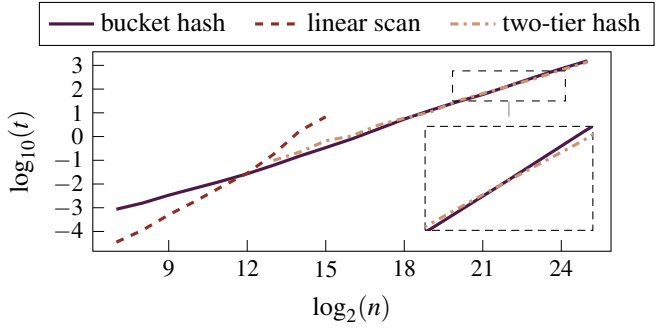
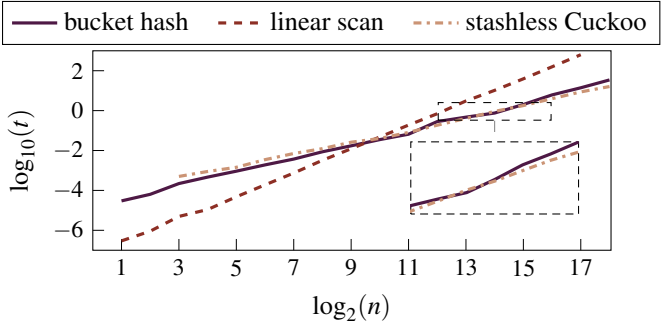
\begin{figure}[t]
    \centering
    \begin{subfigure}[b]{\linewidth}
        \centering
        \begin{tikzpicture}
    \definecolor{darkgray176}{RGB}{176,176,176}
    \definecolor{lightgray204}{RGB}{204,204,204}
    \definecolor{darkslateblue5759121}{RGB}{78,25,69}
    \definecolor{slateblue107110207}{RGB}{141,47,37}
    \definecolor{darkolivegreen9912157}{RGB}{203,148,117}
    \definecolor{darkkhaki181207107}{RGB}{144,146,145}

\begin{axis}[
legend cell align={left},
legend columns=-1, 
legend style={
  fill opacity=0,
  draw opacity=1,
  text opacity=1,
  at={(0.49, 1.25)},
  anchor=north,
  draw=black,
  fill=none,
  /tikz/every even column/.append style={column sep=3mm},
},
log basis x={2},
log basis y={10},
tick align=inside,
tick pos=left,
x grid style={darkgray176},
xlabel={$\log_2(n)$}, %
xmin=68.5935016023227, xmax=62614784.1365567,
xmode=log,
xtick style={color=black},
xtick={8,64,512,4096,32768,262144,2097152,16777216,134217728,1073741824},
xticklabels={
  \(\displaystyle {{3}}\),
  \(\displaystyle {{6}}\),
  \(\displaystyle {{9}}\),
  \(\displaystyle {{12}}\),
  \(\displaystyle {{15}}\),
  \(\displaystyle {{18}}\),
  \(\displaystyle {{21}}\),
  \(\displaystyle {{24}}\),
  \(\displaystyle {{27}}\),
  \(\displaystyle {{30}}\),
},
y grid style={darkgray176},
ylabel={$\log_{10}(t)$}, %
ymin=0.0146793094307909, ymax=3740539.54096949,
ymode=log,
ytick style={color=black},
ytick={0.001,0.01,0.1,1,10,100,1000,10000,100000,1000000,10000000,100000000},
yticklabels={
  \(\displaystyle {{-6}}\),
  \(\displaystyle {{-5}}\),
  \(\displaystyle {{-4}}\),
  \(\displaystyle {{-3}}\),
  \(\displaystyle {{-2}}\),
  \(\displaystyle {{-1}}\),
  \(\displaystyle {{0}}\),
  \(\displaystyle {{1}}\),
  \(\displaystyle {{2}}\),
  \(\displaystyle {{3}}\),
  \(\displaystyle {{4}}\),
  \(\displaystyle {{5}}\),
},
width=\linewidth, 
height=0.5\linewidth,
]
\addplot [very thick, darkslateblue5759121]
table {%
128 0.866058
256 1.57336
512 3.38328
1024 6.63644
2048 13.0328
4096 27.01
8192 61.6965
16384 150.163
32768 342.22
65536 796.668
131072 2069.6
262144 5440.62
524288 12733.4
1048576 28475.2
2097152 59221.8
4194304 136707
8388608 321647
16777216 722129
33554432 1551790
};
\addlegendentry{bucket hash}
\addplot [very thick, slateblue107110207, dashed]
table {%
128 0.035384
256 0.113367
512 0.500643
1024 1.84729
2048 7.46475
4096 28.6402
8192 176.928
16384 1706.63
32768 6811.92
};
\addlegendentry{linear scan}
\addplot [very thick, darkolivegreen9912157, dashdotted]
table {%
8192 98.0003
16384 220.526
32768 661.211
65536 1063.82
131072 3006.45
262144 5953.89
524288 10733.7
1048576 30807.1
2097152 65034.3
4194304 135788
8388608 277282
16777216 639682
33554432 1383170
};
\addlegendentry{two-tier hash}

\coordinate (spypoint) at (axis cs:4194304,136000);

\end{axis}

\node[pin={[pin distance=0.2cm]270:{%
\begin{tikzpicture}[baseline,trim axis left,trim axis right]
    \begin{axis}[
    xtick=\empty, %
    ytick=\empty, %
log basis x={2},
log basis y={10},
tick align=outside,
tick pos=left,
        axis line style={densely dashed},
xmin=2007152, xmax=9088608,
xmode=log,
ymode=log,
ymin= 58221.8, ymax=351647,
ytick style={color=black}, 
tick style={/pgfplots/major tick length=0pt},
height=0.35\linewidth, %
width=0.45\linewidth,
xlabel={},
]
\addplot [very thick, darkslateblue5759121]
table[row sep=crcr] {%
128 0.866058 \\
256 1.57336\\
512 3.38328\\
1024 6.63644\\
2048 13.0328\\
4096 27.01\\
8192 61.6965\\
16384 150.163\\
32768 342.22\\
65536 796.668\\
131072 2069.6\\
262144 5440.62\\
524288 12733.4\\
1048576 28475.2\\
2097152 59221.8\\
4194304 136707\\
8388608 321647\\
16777216 722129\\
33554432 1551790\\
};
\addplot [very thick, darkolivegreen9912157, dashdotted]
table[row sep=crcr] {%
8192 98.0003\\
16384 220.526\\
32768 661.211\\
65536 1063.82\\
131072 3006.45\\
262144 5953.89\\
524288 10733.7\\
1048576 30807.1\\
2097152 65034.3\\
4194304 135788\\
8388608 277282\\
16777216 639682\\
33554432 1383170\\
};
\end{axis}
\end{tikzpicture}
}},draw, dashed, rectangle, minimum width=1.5cm, minimum height=0.4cm] 
at (spypoint) {}; 
\end{tikzpicture}
        \caption{Each hash table is accessed $n$ times.
        Linear scan terminates at $2^{15}$ due to its quadratic growth in running time. 
        Two-tier hash begins at $2^{13}$ as it requires a minimum input size.
        }\label{fig:hash_time:n_t}
    \end{subfigure}
    
    \begin{subfigure}[b]{\linewidth}
        \centering
        \begin{tikzpicture}

    \definecolor{darkgray176}{RGB}{176,176,176}
    \definecolor{lightgray204}{RGB}{204,204,204}
    \definecolor{darkslateblue5759121}{RGB}{78,25,69}
    \definecolor{slateblue107110207}{RGB}{141,47,37}
    \definecolor{darkolivegreen9912157}{RGB}{203,148,117}
    \definecolor{darkkhaki181207107}{RGB}{144,146,145}

\begin{axis}[
legend cell align={left},
legend columns=-1, 
legend style={
  fill opacity=0,
  draw opacity=1,
  text opacity=1,
  at={(0.46, 1.25)},
  anchor=north,
  draw=black,
  fill=none,
  /tikz/every even column/.append style={column sep=3mm},
},
log basis x={2},
log basis y={10},
tick align=inside,
tick pos=left,
x grid style={darkgray176},
xlabel={$\log_2(n)$}, %
xmin=1, xmax=272144,
xmode=log,
xtick style={color=black},
xtick={2,8,32,128,512,2048,8192,32768,131072},
xticklabels={
\(1\),
\(3\),
\(5\),
\(7\),
\(9\),
\(11\),
\(13\),
\(15\),
\(17\),
},
y grid style={darkgray176},
ylabel={$\log_{10}(t)$}, %
ymin=1e-7, ymax=1000,
ymode=log,
ytick style={color=black},
ytick={1e-6, 1e-4, 1e-2, 1, 100},
yticklabels={
  \(\displaystyle {{-6}}\),
  \(\displaystyle {{-4}}\),
  \(\displaystyle {{-2}}\),
  \(\displaystyle {{0}}\),
  \(\displaystyle {{2}}\),
},
width=\linewidth, 
height=0.5\linewidth,
]
\addplot [very thick, darkslateblue5759121]
table {%
2 2.9802082922255895e-05
4 6.369756274973847e-05
8 0.00022057484190416037
16 0.00046434050262419963
32 0.0009206405297280415
64 0.0019014199807511857
128 0.0037215398385039102
256 0.008650575910742657
512 0.01719585167577591
1024 0.035269492763161124
2048 0.0656923232758841
4096 0.2829367960075615
8192 0.4707670359930489
16384 0.7564785060239957
32768 2.0275334549951367
65536 6.148134288028814
131072 13.971388984005898
262144 34.53663087898167
};
\addlegendentry{bucket hash}
\addplot [very thick, slateblue107110207, dashed]
table {%
2 2.9506341095137786e-07
4 9.048069258256468e-07
8 4.966635811130267e-06
16 1.0978809187994521e-05
32 4.788073069620865e-05
64 0.0001897481386707729
128 0.0007426057837153898
256 0.0032356945507840637
512 0.012273848403996805
1024 0.04784640253443892
2048 0.18883663324231748
4096 0.698689906974323
8192 3.05640776033979
16384 10.393375301
32768 39.355034489
65536 157.474238318
131072 634.610902296
};
\addlegendentry{linear scan}
\addplot [very thick, darkolivegreen9912157, dashdotted]
table {%
8 0.000502349
16 0.000889812
32 0.001440685
64 0.003485974
128 0.007071160
256 0.012456863810476879
512 0.026079722147659155
1024 0.04024377358762328
2048 0.08101078656701637
4096 0.18590648075041827
8192 0.4064920929959044
16384 0.9154496989795007
32768 1.7946381730143912
65536 3.983269663003739
131072 8.672826077032367
262144 16.13225515600061
};
\addlegendentry{stashless Cuckoo}
\coordinate (spypoint) at (axis cs:16384,0.9154496989795007);

\end{axis}

\node[pin={[pin distance=0.15cm]270:{%
\begin{tikzpicture}[baseline,trim axis left,trim axis right]
    \begin{axis}[
    xtick=\empty, %
    ytick=\empty, %
log basis x={2},
log basis y={10},
tick align=outside,
tick pos=left,
        axis line style={densely dashed},
xmin=4096, xmax=262144,
xmode=log,
ymode=log,
ymin= 0.15, ymax=35,
ytick style={color=black}, 
tick style={/pgfplots/major tick length=0pt},
height=0.3\linewidth, %
width=0.45\linewidth,
xlabel={},
]
\addplot [very thick, darkslateblue5759121]
table[row sep=crcr]{%
2 2.9802082922255895e-05\\
4 6.369756274973847e-05\\
8 0.00022057484190416037\\
16 0.00046434050262419963\\
32 0.0009206405297280415\\
64 0.0019014199807511857\\
128 0.0037215398385039102\\
256 0.008650575910742657\\
512 0.01719585167577591\\
1024 0.035269492763161124\\
2048 0.0656923232758841\\
4096 0.2829367960075615\\
8192 0.4707670359930489\\
16384 0.7564785060239957\\
32768 2.0275334549951367\\
65536 6.148134288028814\\
131072 13.971388984005898\\
262144 34.53663087898167\\
};
\addplot [very thick, darkolivegreen9912157, dashdotted]
table [row sep=crcr]{%
8 0.000502349\\
16 0.000889812\\
32 0.001440685\\
64 0.003485974\\
128 0.007071160\\
256 0.012456863810476879\\
512 0.026079722147659155\\
1024 0.04024377358762328\\
2048 0.08101078656701637\\
4096 0.18590648075041827\\
8192 0.4064920929959044\\
16384 0.9154496989795007\\
32768 1.7946381730143912\\
65536 3.983269663003739\\
131072 8.672826077032367\\
262144 16.13225515600061\\
};
\end{axis}
\end{tikzpicture}
}},draw, dashed, rectangle, minimum width=1.5cm, minimum height=0.2cm] 
at (spypoint) {}; 
\end{tikzpicture}
        \caption{Each hash table is accessed $128n$ times.}\label{fig:hash_time:n__t}
    \end{subfigure}
    \caption{Overall running time $t$ (in seconds) of different hashing schemes. The block size is $256$B.}\label{fig:hash_time}
\end{figure}

\para{Performance of tailored hash schemes and their optimal use cases.}As shown in~\cref{fig:hash_time}, linear scan performs best when dealing with dozens to hundreds of input data in all cases. 
Bucket hash tables outperform the other methods for medium-sized input data.
While for handling large input data, two-tier hash tables surpass bucket hash tables by $10\%\sim 30\%$ if the number of lookups matches the input data size, as shown in~\cref{fig:hash_time:n_t}.
In contrast, stashless Cuckoo hash tables provide over a $110\%$ speedup compared to bucket hash tables when the number of lookups is 128 times the input data size, as shown in~\cref{fig:hash_time:n__t}.
Note that some lines in the figure appear close to each other because the y-axis is on a logarithmic scale.
The above results are consistent with our analysis in~\cref{tab:ohashes}.

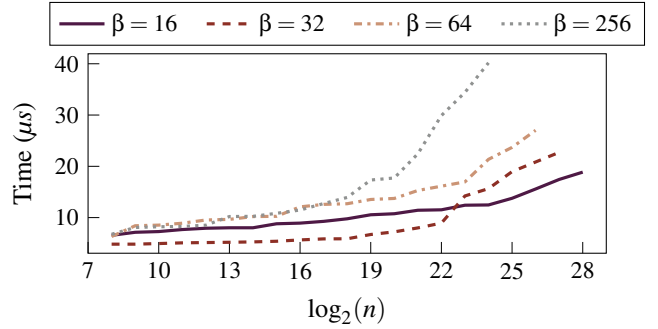
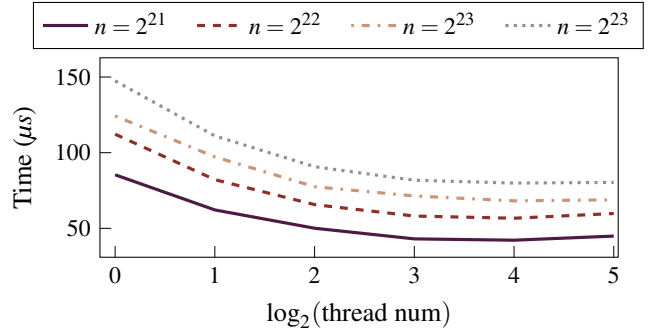
\begin{figure}[t]
    \centering 
    \begin{subfigure}[b]{\linewidth}
        \begin{tikzpicture}
    \definecolor{darkgray176}{RGB}{176,176,176}
    \definecolor{lightgray204}{RGB}{204,204,204}
    \definecolor{darkslateblue5759121}{RGB}{78,25,69}
    \definecolor{slateblue107110207}{RGB}{141,47,37}
    \definecolor{darkolivegreen9912157}{RGB}{203,148,117}
    \definecolor{darkkhaki181207107}{RGB}{144,146,145}

    \begin{axis}[
    legend columns=-1, 
    legend style={
      fill opacity=0,
      draw opacity=1,
      text opacity=1,
      at={(0.5, 1.26)},
      anchor=north,
      draw=black,
      fill=none,
      font=\small,
      /tikz/every even column/.append style={column sep=3mm},
    },
    tick align=inside,
    tick pos=left,
    x grid style={darkgray176},
    xlabel={$\log_2(n)$},
    xmin=128, xmax=536870912,
    log basis x={2},
    xmode=log,
    xtick style={color=black},
    xtick={16,128,1024,8192,65536,524288,4194304,33554432,268435456,2147483648,17179869184},
    xticklabels={
      \(\displaystyle {{4}}\),
      \(\displaystyle {{7}}\),
      \(\displaystyle {{10}}\),
      \(\displaystyle {{13}}\),
      \(\displaystyle {{16}}\),
      \(\displaystyle {{19}}\),
      \(\displaystyle {{22}}\),
      \(\displaystyle {{25}}\),
      \(\displaystyle {{28}}\),
      \(\displaystyle {{31}}\),
      \(\displaystyle {{34}}\)
    },
    y grid style={darkgray176},
    ylabel={Time (\( \mu s\))},
    ymin=3.04978969793215, ymax=41.9504499113666,
    ytick style={color=black},
    height=0.5\linewidth,
    width=\linewidth
    ]
    \addplot [very thick, darkslateblue5759121]
    table {%
    256 6.54296934613287
    512 7.13491490936247
    1024 7.27509473419374
    2048 7.66965327287434
    4096 7.92399594878868
    8192 8.01433294957797
    16384 8.01433710861943
    32768 8.79431510014683
    65536 8.92308990287777
    131072 9.26532652340484
    262144 9.76587277269278
    524288 10.5506561489093
    1048576 10.7517506958033
    2097152 11.4328736801061
    4194304 11.5182075395587
    8388608 12.4160961162873
    16777216 12.4667126923206
    33554432 13.7647100070715
    67108864 15.5769985301644
    134217728 17.4393661862165
    268435456 18.8656661845519
    };
    \addlegendentry{$\beta=16$}
    \addplot [very thick, slateblue107110207, dashed]
    table {%
    256 4.81800152581554
    512 4.82024929198133
    1024 4.93820040137205
    2048 5.07463535852943
    4096 5.13224632504641
    8192 5.18655235293153
    16384 5.26023646486351
    32768 5.38462081145497
    65536 5.64624657855176
    131072 5.84820908768673
    262144 5.86974466056358
    524288 6.70318015291715
    1048576 7.21722902202737
    2097152 7.98958930397359
    4194304 8.92496210670377
    8388608 14.2211073001618
    16777216 15.6476887732144
    33554432 18.9670538678168
    67108864 20.8441661535205
    134217728 22.720069
    };
    \addlegendentry{$\beta=32$}
    \addplot [very thick, darkolivegreen9912157, dashdotted]
    table {%
    256 6.30746284909801
    512 8.41064857155516
    1024 8.53380089319937
    2048 8.90345374133061
    4096 9.53252568195701
    8192 9.65087525939268
    16384 10.1968837928901
    32768 10.249757124502
    65536 12.093091235499
    131072 12.5777363662594
    262144 12.7042406186295
    524288 13.5346028089511
    1048576 13.7357775116254
    2097152 15.2866802520663
    8388608 16.9682150797935
    16777216 21.3618580957523
    33554432 23.69442532437
    67108864 27.0235891230254
    };
    \addlegendentry{$\beta=64$}
    \addplot [very thick, darkkhaki181207107, dotted]
    table {%
    256 6.75039786688945
    512 8.07513235479007
    1024 8.21375396675705
    2048 8.27623842702962
    4096 8.51879712100402
    8192 10.20671233666
    16384 10.2421589705922
    32768 10.8349959297982
    65536 11.4755336918885
    131072 12.7382555201894
    262144 13.9294491465769
    524288 17.3289787154252
    1048576 17.7255055284764
    2097152 22.4439434347024
    4194304 29.9011968245277
    8388608 34.4228516284822
    16777216 40.1822380834832
    };
    \addlegendentry{$\beta=256$}
    
    \end{axis}
\end{tikzpicture}
        \caption{Performance of \system under various input data sizes and data block sizes $\beta$.}\label{fig:h2o2ram:n}
    \end{subfigure}
    
    \begin{subfigure}[b]{\linewidth}
        \begin{tikzpicture}
    \definecolor{darkgray176}{RGB}{176,176,176}
    \definecolor{lightgray204}{RGB}{204,204,204}
    \definecolor{darkslateblue5759121}{RGB}{78,25,69}
    \definecolor{slateblue107110207}{RGB}{141,47,37}
    \definecolor{darkolivegreen9912157}{RGB}{203,148,117}
    \definecolor{darkkhaki181207107}{RGB}{144,146,145}

    \begin{axis}[
    legend columns=-1, 
    legend style={
      fill opacity=0,
      draw opacity=1,
      text opacity=1,
      at={(0.46, 1.28)},
      anchor=north,
      draw=black,
      fill=none,
      font=\small,
      /tikz/every even column/.append style={column sep=3mm},
    },
    tick align=inside,
    tick pos=left,
    x grid style={darkgray176},
    xlabel={$\log_2($thread num$)$},
    xmin=0.9, xmax=33,
    log basis x={2},
    xmode=log,
    xtick style={color=black},
    xtick={1,2,4,8,16,32},
    xticklabels={0,1,2,3,4,5},
    y grid style={darkgray176},
    ylabel={Time (\( \mu s\))},
    ymin=30.8986299426499, ymax=162.727291529553,
    ytick style={color=black},
    height=0.5\linewidth,
    width=\linewidth
    ]
    \addplot [very thick, darkslateblue5759121]
    table {%
    1 85.3941710200479
    2 62.1711631369481
    4 50.073922009064
    8 43.038956399899
    16 42.163569105691
    32 44.8878868694047
    };
    \addlegendentry{$n=2^{21}$}
    \addplot [very thick, slateblue107110207, dashed]
    table {%
    1 112.222935187339
    2 82.1745746960589
    4 65.6761912863346
    8 58.1304475579247
    16 56.6874847614529
    32 59.8023936490555
    };
    \addlegendentry{$n=2^{22}$}
    \addplot [very thick, darkolivegreen9912157, dash pattern=on 1pt off 3pt on 3pt off 3pt]
    table {%
    1 124.288817032581
    2 97.2370239087361
    4 77.5037404017465
    8 71.3984389775768
    16 68.1667839659561
    32 68.8457032569645
    };
    \addlegendentry{$n=2^{23}$}
    \addplot [very thick, darkkhaki181207107, dotted]
    table {%
    1 147.462352366512
    2 111.111527783689
    4 90.7403209067653
    8 81.8030014392152
    16 79.9029945396795
    32 80.3644761669664
    };
    \addlegendentry{$n=2^{23}$}
    
    \end{axis}
\end{tikzpicture}
        \caption{Performance of \system under different degrees of parallelization and input data sizes.}\label{fig:h2o2ram:thread}
    \end{subfigure}
    \caption{Amortized access time of \system across different problem sizes. 
    }\label{fig:h2o2ram}
\end{figure}

\para{Doubly oblivious RAM.}We then evaluate the performance of \system in different scenarios. 
We generate $n$ data blocks with keys from $0$ to $n-1$ and values of random bits.
These blocks are then randomly shuffled to permute the data, upon which \system is built.
We also amortize the building time across $n$ data accesses. 
As shown in~\cref{fig:h2o2ram:n}, the amortized access time for \system increases polylogarithmically with its capacity $n$, and increases approximately linearly with the data block size. 
\Cref{fig:h2o2ram:thread} shows that parallelization effectively improves the access efficiency of \system with performance gains observed up to $16$ threads.
Note that our implementation may achieve only suboptimal parallelization due to our limited expertise in this area, suggesting potential for further performance improvements.

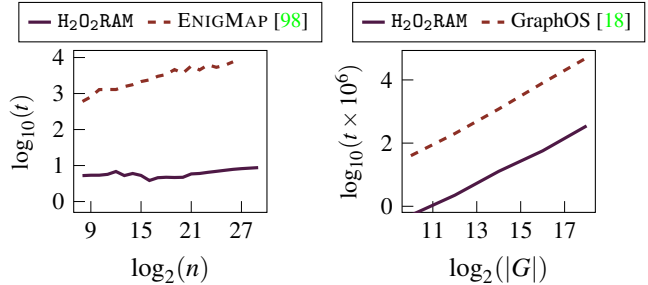
\begin{figure}[th]
    \centering
    \begin{subfigure}[t]{0.49\linewidth}
        \begin{tikzpicture}
    \definecolor{darkgray176}{RGB}{176,176,176}
    \definecolor{lightgray204}{RGB}{204,204,204}
    \definecolor{darkslateblue5759121}{RGB}{78,25,69}
    \definecolor{slateblue107110207}{RGB}{141,47,37}
    \definecolor{darkolivegreen9912157}{RGB}{203,148,117}
    \definecolor{darkkhaki181207107}{RGB}{144,146,145}

    \begin{axis}[
    legend columns=-1, 
    legend style={
      fill opacity=0,
      draw opacity=1,
      text opacity=1,
      at={(0.5, 1.3)},
      anchor=north,
      draw=black,
      fill=none,
      font=\footnotesize,
      /tikz/every even column/.append style={column sep=1mm},
    },
    legend image post style={xscale=0.5}, 
    tick align=inside,
    tick pos=left,
    x grid style={darkgray176},
    xlabel={$\log_2(n)$},
    xmin=123.63985010238, xmax=1111607247.64866,
    log basis x={2},
    xmode=log,
    xtick style={color=black},
    xtick={512,32768,2097152,134217728},
    xticklabels={
      \(\displaystyle {9}\),
      \(\displaystyle {15}\),
      \(\displaystyle {21}\),
      \(\displaystyle {27}\),
    },
    ylabel={\small $\log_{10}(t)$},
    ymin=0.5, ymax=15704.9431740687,
    log basis y={10},
    ymode=log,
    ytick style={color=black},
    ytick={0.0001,0.001,0.01,0.1,1,10,100,1000,10000,100000,1000000},
    yticklabels={
      \(\displaystyle {-4}\),
      \(\displaystyle {-3}\),
      \(\displaystyle {-2}\),
      \(\displaystyle {-1}\),
      \(\displaystyle {0}\),
      \(\displaystyle {1}\),
      \(\displaystyle {2}\),
      \(\displaystyle {3}\),
      \(\displaystyle {4}\),
      \(\displaystyle {5}\),
      \(\displaystyle {6}\)
    },
    yminorticks=true,
    height=0.9\linewidth,
    width=\linewidth
    ]
    \addplot [very thick, darkslateblue5759121]
    table {%
    256 5.25237599001056
    512 5.38710308898193
    1024 5.38896562128515
    2048 5.69155965247072
    4096 6.83637622008367
    8192 5.26530541075469
    16384 6.01788390294392
    32768 5.31092923949927
    65536 3.84384922706242
    131072 4.59107120942548
    262144 4.72641092308379
    524288 4.65467179644463
    1048576 4.71582155814332
    2097152 5.84938487246944
    4194304 6.0221738162336
    8388608 6.48354983961985
    16777216 6.90393152832924
    33554432 7.36520369810228
    67108864 7.84246811365991
    134217728 8.19841794541531
    268435456 8.47717545422906
    536870912 8.75443601110366
    };
    \addlegendentry{\system}
    \addplot [very thick, slateblue107110207, dashed]
    table {%
    256 603.018
    512 807.782
    1024 1288.55
    2048 1296.88
    4096 1292.71
    8192 1560.18
    16384 1760.38
    32768 2153.15
    65536 2397.4
    131072 3008.9
    262144 3378.59
    524288 4623.92
    1048576 3670.67
    2097152 5831.9
    4194304 4531.52
    8388608 6268.67
    16777216 5351.69
    33554432 6089.26
    67108864 7607.01
    };
    \addlegendentry{\textsc{EnigMap}~\cite{tinoco2023enigmap}}
    \end{axis}
\end{tikzpicture}
        \caption{\small \textsf{Get} operation of a map, where $n$ denotes its capacity.
        }\label{fig:cmp:map}
    \end{subfigure}
    \hfill
    \begin{subfigure}[t]{0.49\linewidth}
        \begin{tikzpicture}
    \definecolor{darkgray176}{RGB}{176,176,176}
    \definecolor{lightgray204}{RGB}{204,204,204}
    \definecolor{darkslateblue5759121}{RGB}{78,25,69}
    \definecolor{slateblue107110207}{RGB}{141,47,37}
    \definecolor{darkolivegreen9912157}{RGB}{203,148,117}
    \definecolor{darkkhaki181207107}{RGB}{144,146,145}

    \begin{axis}[
    legend columns=-1, 
    legend style={
      fill opacity=0,
      draw opacity=1,
      text opacity=1,
      at={(0.5, 1.3)},
      anchor=north,
      draw=black,
      fill=none,
      font=\footnotesize,
      /tikz/every even column/.append style={column sep=1mm},
    },
    legend image post style={xscale=0.5}, 
    tick align=inside,
    tick pos=left,
    x grid style={darkgray176},
    xlabel={$\log_2(|G|)$},
    xmin=776.046882053324, xmax=345901.081761649,
    log basis x={2},
    xmode=log,
    xtick style={color=black},
    xtick={128,512,2048,8192,32768,131072,524288},
    xticklabels={
      \(\displaystyle {7}\),
      \(\displaystyle {9}\),
      \(\displaystyle {11}\),
      \(\displaystyle {13}\),
      \(\displaystyle {15}\),
      \(\displaystyle {17}\),
      \(\displaystyle {19}\),
    },
    ylabel={\small $\log_{10}(t\times 10^{6})$},
    ymin=0.641070596018104, ymax=85491.9926792336,
    log basis y={10},
    ymode=log,
    ytick style={color=black},
    ytick={1,100,10000},
    yticklabels={
      0,
      2,
      4,
    },
    yminorticks=true,
    height=0.9\linewidth,
    width=\linewidth
    ]
    \addplot [very thick, darkslateblue5759121]
    table {%
    1024 0.52196574001573
    4096 2.23795651295222
    16384 12.7662738440558
    65536 56.8635256480193
    262144 349.892674535979
    };
    \addlegendentry{\system}
    \addplot [very thick, slateblue107110207, dashed]
    table {%
    1024 40
    4096 200
    16384 1200
    65536 8000
    262144 50000
    };
    \addlegendentry{GraphOS~\cite{chamani2023graphos}}
    \end{axis}
\end{tikzpicture}
        \caption{\small Single-source shortest path computation. 
        The graph size $G$ equals to the sum of the vertices count $|V|$ and the edge count $|E|$ with $|E|=4|V|$.
        }\label{fig:cmp:sssp}
    \end{subfigure}
    \caption{Comparison of \system and \textsc{EnigMap}~\cite{tinoco2023enigmap} / GraphOS~\cite{chamani2023graphos} in terms of computation time $t$ in $\mu$s.
    }\label{fig:cmp}
\end{figure}
\para{Doubly oblivious tasks.}We also evaluate the performance of \system when applied in the computations of single-source shortest paths and the key-value map data structure, and draw a comparison with GraphOS~\cite{chamani2023graphos} and \textsc{EnigMap}~\cite{tinoco2023enigmap}, resp. 
As shown in~\cref{fig:cmp:map}, our solution reduces the \textsf{Get} operation time of a map by a factor of $100\times$ to $997\times$ compared to \textsc{EnigMap}~\cite{tinoco2023enigmap}. 
As for single-source shortest path computation, \system is in computing single-source shortest paths with total computation time improvements ranging from $76.6\times$ to $142.3\times$, reducing the computation task ($|G|=2^{18}$) required by GraphOS from over half a day ($\sim 13$ hours) to less than 6 minutes ($349$ s).
Note that our advantage over GraphOS is less pronounced compared to~\cref{fig:cmp:map} due to the relatively ``small'' graph sizes in our experiments. 
For larger graphs, GraphOS would require several days or even weeks to complete the computations.

In addition to better running time, \system also enjoys better memory space overhead.
For instance, with $2^{22}$/$2^{23}$ blocks of 16 data blocks (a total of $64$/$128$ MB of data), \system requires $0.82$/$1.63$ GB of memory, whereas \textsc{EnigMap} consumes $4.75$/$72.3$ GB memory (\ie, \system reduces memory overhead by a factor of $5.79$ to $44.36$).
Note that \textsc{EnigMap} experiences such an ``exploded'' memory consumption issue due to its packing strategies, which are designed for better locality, an inherent advantage enjoyed by \system.

\section{Related Work}\label{sec:related}
\para{Oblivious RAM.}Since the seminal work in~\cite{goldreich1987towards, goldreich1996software}, ORAM and its variants have attracted widespread interest in many applications, including but not limited to cloud computing~\cite{chamani2023graphos, bindschaedler2015practicing, fletcher2015freecursive, gentry2015private, OstrovskyS97, StefanovS13, stefanov2011towards, williams2012privatefs}, multi-party computation~\cite{ liu2015oblivm, lu2013distributed, NobleHO24, wang2014scoram_linear, zahur2016revisiting}, and secure processor designs~\cite{fletcher2012secure, LiuHMHTS15, maas2013phantom, RenYFDD13}. 
From a theoretical perspective, an $\Omega(\log n)$ lower bound shown in the very first study has been extended to many different settings~\cite{larsen2018yes, boyle2016there, cash2020lower, jacob2019lower, weiss2021there, larsen2020lower, komargodski2021logarithmic, larsen2020lower}. 
While it took decades to develop solutions~\cite{asharov2022optorama, asharov2022optimal} that meet this lower bound, most of the efforts~\cite{PatelP0Y18PanORAMa, pinkas2010oblivious, chan2017obliviousCuckoo, stefanov2018path, wang2015circuit, kushilevitz2012security} over this span were based on the hierarchical framework.
However,  a solution that is both asymptotically optimal and practically efficient, without relying on additional assumptions, remains unknown.
On the other hand, Path ORAM~\cite{stefanov2018path}, a simple yet powerful tree-based design, has opened up numerous practical applications for ORAM. 
Afterwards, a variety of optimized variants~\cite{mishra2018oblix, dauterman2021snoopy, chamani2023graphos, bindschaedler2015practicing, NobleHO24} have been developed for different application scenarios.
The most relevant to us among these is ORAM with secure hardware.
\citet{shi2020path} proposed a novel doubly oblivious heap and other data structures. 
ZeroTrace~\cite{sasy2017zerotrace} offers constructions based on Path ORAM and Circuit ORAM~\cite{wang2015circuit}, which is soon outperformed by \textsc{Oblix}~\cite{mishra2018oblix}. 
A very recent work, GraphOS~\cite{chamani2023graphos}, further optimizes this design and offers several doubly oblivious graph algorithms. 
\textsc{ENIGMAP}~\cite{tinoco2023enigmap} uses external-memory algorithms to optimize oblivious accesses. 
Obliviate~\cite{ahmad2018obliviate}, MOSE~\cite{hoang2020mose}, POSUP~\cite{hoang2019hardware_posup}, Shroud~\cite{lorch2013shroud}, and Snoopy~\cite{dauterman2021snoopy} adopt different optimizations for oblivious storage systems. 
\textsc{Phantom}~\cite{maas2013phantom}, GhostRider~\cite{LiuHMHTS15}, and Tiny ORAM~\cite{fletcher2015low} use FPGA as the backend trusted hardware. 

\para{Oblivious primitives.}Our work also heavily relies on various oblivious operations and primitives. 
T-SGX~\cite{shih2017t} focuses on eradicating controlled-channel attacks. 
ObliVM~\cite{liu2015oblivm}, though relying on multiple noncolluding servers, proposed several operators for general oblivious execution.
Obfuscuro~\cite{ahmad2019obfuscuro}, Klotski~\cite{zhang2020klotski}, and Obelix~\cite{WichelmannRPE24} stand for the most advanced approaches to oblivious execution of arbitrary code. 
\citet{sinha2017compiler} implement an efficient compiler to enforce page-access obliviousness for a type and memory-safe languages. 
ObliCheck~\cite{son2021oblicheck} can efficiently verify whether an algorithm is indeed oblivious. 
In addition, there have been significant recent advancements in oblivious sorting, shuffling, and compaction algorithms~\cite{sasy2022fastocompact, Sasy0G23Waks}.
In particular, Bitonic sorting/shuffling has long been considered the most concretely efficient algorithm, but it has been consistently outperformed by~\cite{sasy2022fastocompact} and its subsequent follow-up~\cite{Sasy0G23Waks}. 
However, such designs rely on an expensive offline preprocessing stage that generates some pseudorandomness, which makes it inappropriate for our design, as \system frequently invokes osort/oshuffling.

\para{Relaxed oblivious designs.}Designs that trade full obliviousness for better performance are also prevalent.
Examples include straightforward relaxations such as page-level obliviousness~\cite{sinha2017compiler, ShindeCNS16page1, TopleS17page2, RachidRM20page3} and the existence of an inherently oblivious private cache~\cite{EskandarianZ19, NagarajanSBT19}.
\citet{GrubbsKLBL0R20} and \citet{MaiyyaVAAK23} introduce relaxed notions of obliviousness in the context of key-value stores. 
In addition, differentially oblivious designs~\cite{LeqianSWAT, ChanCMS19, QiuKMNK23, WuNHL23, Qin22Adore, GordonKLX22}, which bring the ideology of differential privacy, provide well-structured security notions.

\section{Conclusion}\label{sec:conclusion}
In this work, we take a step forward in the practical application of hierarchical ORAM with Trusted Execution Environments,  further underscoring the potential of the hierarchical ORAM framework.
Our design, \system, offers a general and efficient approach for executing algorithms that require the protection of access patterns. 
Moreover, the components we introduced, such as the oblivious bipartite matching and the oblivious stashless Cuckoo hash table, are of independent interest.
We have also implemented and open-sourced \system, and conducted empirical evaluations in various scenarios to show its concrete efficiency. 
The results show that \system surpasses state-of-the-art designs by up to $\sim 10^3\times$ in running time and by $44\times$ in space consumption.

\section*{Acknowledgements}
We appreciate the detailed and valuable feedback provided by our anonymous reviewers. 
We thank Qingling Feng and Dian Chen for their discussions and for their spare time in running experiments. 
This work was supported in part by the Research Grants Council of Hong Kong under Grants CityU 11218322, 11219524, R6021-20F, R1012-21, RFS2122-1S04, C2004-21G, C1029-22G, C6015-23G, and N\_CityU139/21 and in part by the Innovation and Technology Commission of Hong Kong (ITC) under Mainland-Hong Kong Joint Funding Scheme (MHKJFS) under Grant MHP/135/23. 
This work was also supported by the InnoHK initiative, the Government of the HKSAR, and the Laboratory for AI-Powered Financial Technologies (AIFT).

\section*{Ethics Considerations}
Our research adheres to stringent ethical standards to ensure the integrity and societal impact of the work presented. 
The core ethical considerations in this study include the protection of data privacy in cloud computing, and the broader implications of the technology we develop. 
We conducted our evaluation experiments using synthetic (randomly generated) data.
Indeed, our objective is to transform programs into forms where their behavior is (pseudo-)independent of the input data, which may contain highly sensitive private information.
Consequently, our work contributes to safeguarding user privacy in a broad spectrum of applications, including but not limited to searchable encryption, private contact discovery in end-to-end messaging, key transparency, and anonymous broadcasting/subscription platforms. 
However, we emphasize the following two major ethical issues that must be considered when applying our techniques:
\begin{enumerate}
    \item Our approach relies heavily on redundant computations to safeguard data privacy, potentially resulting in the waste use of computational resources and, more critically, increased energy consumption.
    For instance, we observed that our server was operating at nearly full power during the experiments, while non-privacy-preserving solutions were merely accessing random addresses.
    Therefore, it is essential to evaluate whether such a high level of protection is necessary when designing a privacy-preserving application. 
    Blindly applying such techniques will definitely lead to significant energy/resource waste and environmental pollution.
    \item One important application of our work is to ensure anonymity.
    Maintaining online anonymity is a fundamental right for everyone.
    However, anonymity can occasionally be associated with hate speech or criminal activities. 
    Therefore, those using our technology should also consider how to prevent inappropriate dissemination of offensive content or information.
\end{enumerate}
In short, our work primarily focuses on privacy-preserving applications in cloud computing; however, special attention must still be given to the two aforementioned issues when applying this technology.

To ensure that our research is both transparent and reproducible, all code, data, and experimental results will be made available to the community, subject to the constraints of the Open Science Policy discussed below.

\section*{Open Science Compliance}
In compliance with the Open Science Policy adopted by USENIX and other leading research communities, we commit to sharing our research artifacts in a way that promotes transparency, reproducibility, and accessibility. 
We share our code on \codes with clear instructions on reproducing the experiments.

{%
\bibliographystyle{plainnat}
\bibliography{ref}

\begin{thebibliography}{111}
\providecommand{\natexlab}[1]{#1}
\providecommand{\url}[1]{\texttt{#1}}
\expandafter\ifx\csname urlstyle\endcsname\relax
  \providecommand{\doi}[1]{doi: #1}\else
  \providecommand{\doi}{doi: \begingroup \urlstyle{rm}\Url}\fi

\bibitem[Afonso et~al.()Afonso, Sixiang, and Elaine]{enigmap_code}
Tinoco Afonso, Gao Sixiang, and Shi Elaine.
\newblock {EnigMap}: Oblivious data structure library.
\newblock URL \url{https://github.com/odslib/EnigMap}.

\bibitem[Ahmad et~al.(2018)Ahmad, Kim, Sarfaraz, and Lee]{ahmad2018obliviate}
Adil Ahmad, Kyungtae Kim, Muhammad~Ihsanulhaq Sarfaraz, and Byoungyoung Lee.
\newblock Obliviate: A data oblivious filesystem for {Intel} {SGX}.
\newblock In \emph{NDSS}, 2018.

\bibitem[Ahmad et~al.(2019)Ahmad, Joe, Xiao, Zhang, Shin, and Lee]{ahmad2019obfuscuro}
Adil Ahmad, Byunggill Joe, Yuan Xiao, Yinqian Zhang, Insik Shin, and Byoungyoung Lee.
\newblock Obfuscuro: A commodity obfuscation engine on {Intel} {SGX}.
\newblock In \emph{NDSS}, 2019.

\bibitem[AMD()]{AMDSEV}
AMD.
\newblock {AMD SEV-SNP}: Strengthening {VM} isolation with integrity protection and more.
\newblock URL \url{https://www.amd.com/content/dam/amd/en/documents/epyc-business-docs/white-papers/SEV-SNP-strengthening-vm-isolation-with-integrity-protection-and-more.pdf}.

\bibitem[Asharov et~al.(2020)Asharov, Chan, Nayak, Pass, Ren, and Shi]{asharov2020bucket}
Gilad Asharov, TH~Hubert Chan, Kartik Nayak, Rafael Pass, Ling Ren, and Elaine Shi.
\newblock Bucket oblivious sort: An extremely simple oblivious sort.
\newblock In \emph{SOSA}, 2020.

\bibitem[Asharov et~al.(2022{\natexlab{a}})Asharov, Komargodski, Lin, Nayak, Peserico, and Shi]{asharov2022optorama}
Gilad Asharov, Ilan Komargodski, Wei-Kai Lin, Kartik Nayak, Enoch Peserico, and Elaine Shi.
\newblock {OptORAMa}: Optimal oblivious {RAM}.
\newblock \emph{JACM}, 70\penalty0 (1):\penalty0 1--70, 2022{\natexlab{a}}.

\bibitem[Asharov et~al.(2022{\natexlab{b}})Asharov, Komargodski, Lin, Peserico, and Shi]{asharov2022optimal}
Gilad Asharov, Ilan Komargodski, Wei-Kai Lin, Enoch Peserico, and Elaine Shi.
\newblock Optimal oblivious parallel {RAM}.
\newblock In \emph{SODA}, 2022{\natexlab{b}}.

\bibitem[Asharov et~al.(2023)Asharov, Komargodski, and Michelson]{asharov2023futorama}
Gilad Asharov, Ilan Komargodski, and Yehuda Michelson.
\newblock {FutORAMa}: A concretely efficient hierarchical oblivious {RAM}.
\newblock In \emph{CCS}, 2023.

\bibitem[Bindschaedler et~al.(2015)Bindschaedler, Naveed, Pan, Wang, and Huang]{bindschaedler2015practicing}
Vincent Bindschaedler, Muhammad Naveed, Xiaorui Pan, XiaoFeng Wang, and Yan Huang.
\newblock Practicing oblivious access on cloud storage: the gap, the fallacy, and the new way forward.
\newblock In \emph{CCS}, 2015.

\bibitem[Board()]{openmp}
OpenMP Architecture~Review Board.
\newblock The {OpenMP} {API} specification for parallel programming.
\newblock URL \url{https://www.openmp.org/}.

\bibitem[Boyle and Naor(2016)]{boyle2016there}
Elette Boyle and Moni Naor.
\newblock Is there an oblivious {RAM} lower bound?
\newblock In \emph{ITCS}, 2016.

\bibitem[Brasser et~al.(2017)Brasser, Müller, Dmitrienko, Kostiainen, Capkun, and Sadeghi]{brasser2017software}
Ferdinand Brasser, Urs Müller, Alexandra Dmitrienko, Kari Kostiainen, Srdjan Capkun, and Ahmad-Reza Sadeghi.
\newblock Software grand exposure: {SGX} cache attacks are practical.
\newblock In \emph{WOOT}, 2017.

\bibitem[Bulck et~al.(2017)Bulck, Weichbrodt, Kapitza, Piessens, and Strackx]{Bulck2017TellingYS}
Jo~Van Bulck, Nico Weichbrodt, R{\"{u}}diger Kapitza, Frank Piessens, and Raoul Strackx.
\newblock Telling your secrets without page faults: Stealthy page table-based attacks on enclaved execution.
\newblock In \emph{{USENIX} Security}, 2017.

\bibitem[Bulck et~al.(2018)Bulck, Minkin, Weisse, Genkin, Kasikci, Piessens, Silberstein, Wenisch, Yarom, and Strackx]{BulckMWGKPSWYS18}
Jo~Van Bulck, Marina Minkin, Ofir Weisse, Daniel Genkin, Baris Kasikci, Frank Piessens, Mark Silberstein, Thomas~F. Wenisch, Yuval Yarom, and Raoul Strackx.
\newblock Foreshadow: Extracting the keys to the intel {SGX} kingdom with transient out-of-order execution.
\newblock In \emph{{USENIX} Security}, 2018.

\bibitem[Cash et~al.(2015)Cash, Grubbs, Perry, and Ristenpart]{Cash2015LeakageAbuseAA}
David Cash, Paul Grubbs, Jason Perry, and Thomas Ristenpart.
\newblock Leakage-abuse attacks against searchable encryption.
\newblock In \emph{CCS}, 2015.

\bibitem[Cash et~al.(2020)Cash, Drucker, and Hoover]{cash2020lower}
David Cash, Andrew Drucker, and Alexander Hoover.
\newblock A lower bound for one-round oblivious {RAM}.
\newblock In \emph{TCC}, 2020.

\bibitem[Chakraborti and Sion(2019)]{ChakrabortiS19}
Anrin Chakraborti and Radu Sion.
\newblock Concuroram: High-throughput stateless parallel multi-client {ORAM}.
\newblock In \emph{{NDSS}}, 2019.

\bibitem[Chamani et~al.(2023)Chamani, Demertzis, Papadopoulos, Papamanthou, and Jalili]{chamani2023graphos}
Javad~Ghareh Chamani, Ioannis Demertzis, Dimitrios Papadopoulos, Charalampos Papamanthou, and Rasool Jalili.
\newblock {GraphOS}: Towards oblivious graph processing.
\newblock In \emph{VLDB}, 2023.

\bibitem[Chan et~al.(2017)Chan, Guo, Lin, and Shi]{chan2017obliviousCuckoo}
T-H~Hubert Chan, Yue Guo, Wei-Kai Lin, and Elaine Shi.
\newblock Oblivious hashing revisited, and applications to asymptotically efficient oram and opram.
\newblock In \emph{ASIACRYPT}, 2017.

\bibitem[Chan et~al.(2019)Chan, Chung, Maggs, and Shi]{ChanCMS19}
T.{-}H.~Hubert Chan, Kai{-}Min Chung, Bruce~M. Maggs, and Elaine Shi.
\newblock Foundations of differentially oblivious algorithms.
\newblock In \emph{{SODA}}, pages 2448--2467. {SIAM}, 2019.

\bibitem[Chawla(2009)]{balls_into_bins}
Shuchi Chawla.
\newblock Lecture 7: Randomized load balancing and hashing, 2009.

\bibitem[Chen et~al.(2020)Chen, Chen, Xiao, Zhang, Lin, and Lai]{ChenCXZLL20}
Guoxing Chen, Sanchuan Chen, Yuan Xiao, Yinqian Zhang, Zhiqiang Lin, and Ten{-}Hwang Lai.
\newblock Sgxpectre: Stealing intel secrets from {SGX} enclaves via speculative execution.
\newblock \emph{{IEEE} S\&P}, 18\penalty0 (3):\penalty0 28--37, 2020.

\bibitem[Cloud()]{AlibabaTEE}
Alibaba Cloud.
\newblock {TEE}-based confidential computing.
\newblock URL \url{https://www.alibabacloud.com/help/en/ack/ack-managed-and-ack-dedicated/user-guide/tee-based-confidential-computing}.

\bibitem[Dauterman et~al.(2021)Dauterman, Fang, Demertzis, Crooks, and Popa]{dauterman2021snoopy}
Emma Dauterman, Vivian Fang, Ioannis Demertzis, Natacha Crooks, and Raluca~Ada Popa.
\newblock Snoopy: Surpassing the scalability bottleneck of oblivious storage.
\newblock In \emph{SOSP}, 2021.

\bibitem[Dittmer and Ostrovsky(2020)]{DittmerO20}
Samuel Dittmer and Rafail Ostrovsky.
\newblock Oblivious tight compaction in {O(n)} time with smaller constant.
\newblock In \emph{SCN}, 2020.

\bibitem[Doerner and Shelat(2017)]{doerner2017scaling_linear}
Jack Doerner and Abhi Shelat.
\newblock Scaling {ORAM} for secure computation.
\newblock In \emph{CCS}, 2017.

\bibitem[Eskandarian and Zaharia(2019)]{EskandarianZ19}
Saba Eskandarian and Matei Zaharia.
\newblock {ObliDB}: Oblivious query processing for secure databases.
\newblock \emph{{VLDB}}, 13\penalty0 (2):\penalty0 169--183, 2019.

\bibitem[Fletcher et~al.(2012)Fletcher, Dijk, and Devadas]{fletcher2012secure}
Christopher~W Fletcher, Marten~van Dijk, and Srinivas Devadas.
\newblock A secure processor architecture for encrypted computation on untrusted programs.
\newblock In \emph{STC}, 2012.

\bibitem[Fletcher et~al.(2015{\natexlab{a}})Fletcher, Ren, Kwon, Van~Dijk, and Devadas]{fletcher2015freecursive}
Christopher~W Fletcher, Ling Ren, Albert Kwon, Marten Van~Dijk, and Srinivas Devadas.
\newblock Freecursive {ORAM}: Nearly free recursion and integrity verification for position-based oblivious {RAM}.
\newblock In \emph{ASPLOS}, 2015{\natexlab{a}}.

\bibitem[Fletcher et~al.(2015{\natexlab{b}})Fletcher, Ren, Kwon, Van~Dijk, Stefanov, Serpanos, and Devadas]{fletcher2015low}
Christopher~W Fletcher, Ling Ren, Albert Kwon, Marten Van~Dijk, Emil Stefanov, Dimitrios Serpanos, and Srinivas Devadas.
\newblock A low-latency, low-area hardware oblivious {RAM} controller.
\newblock In \emph{FCCM}, 2015{\natexlab{b}}.

\bibitem[Fotakis et~al.(2005)Fotakis, Pagh, Sanders, and Spirakis]{FotakisPSS05}
Dimitris Fotakis, Rasmus Pagh, Peter Sanders, and Paul~G. Spirakis.
\newblock Space efficient hash tables with worst case constant access time.
\newblock \emph{TOCS}, 38\penalty0 (2):\penalty0 229--248, 2005.

\bibitem[Gentry et~al.(2015)Gentry, Halevi, Jutla, and Raykova]{gentry2015private}
Craig Gentry, Shai Halevi, Charanjit Jutla, and Mariana Raykova.
\newblock Private database access with {HE-over-ORAM} architecture.
\newblock In \emph{ACNS}, 2015.

\bibitem[Goldreich(1987)]{goldreich1987towards}
Oded Goldreich.
\newblock Towards a theory of software protection and simulation by oblivious {RAMs}.
\newblock In \emph{STOC}, 1987.

\bibitem[Goldreich and Ostrovsky(1996)]{goldreich1996software}
Oded Goldreich and Rafail Ostrovsky.
\newblock Software protection and simulation on oblivious {RAMs}.
\newblock \emph{JACM}, 43\penalty0 (3):\penalty0 431--473, 1996.

\bibitem[Goodrich and Mitzenmacher(2011)]{goodrich2011cuckoo}
Michael~T Goodrich and Michael Mitzenmacher.
\newblock Privacy-preserving access of outsourced data via oblivious {RAM} simulation.
\newblock In \emph{ICALP}, 2011.

\bibitem[Google({\natexlab{a}})]{GoogleTEE}
Google.
\newblock Confidential computing, {\natexlab{a}}.
\newblock URL \url{https://cloud.google.com/security/products/confidential-computing?hl=en}.

\bibitem[Google({\natexlab{b}})]{googleBenchmark}
Google.
\newblock {Google Benchmark}, {\natexlab{b}}.
\newblock URL \url{https://github.com/google/benchmark}.

\bibitem[Gordon et~al.(2022)Gordon, Katz, Liang, and Xu]{GordonKLX22}
S.~Dov Gordon, Jonathan Katz, Mingyu Liang, and Jiayu Xu.
\newblock Spreading the privacy blanket: Differentially oblivious shuffling for differential privacy.
\newblock In \emph{ACNS}, 2022.

\bibitem[Grubbs et~al.(2016)Grubbs, McPherson, Naveed, Ristenpart, and Shmatikov]{Grubbs2016BreakingWA}
Paul Grubbs, Richard McPherson, Muhammad Naveed, Thomas Ristenpart, and Vitaly Shmatikov.
\newblock Breaking web applications built on top of encrypted data.
\newblock In \emph{CCS}, 2016.

\bibitem[Grubbs et~al.(2020)Grubbs, Khandelwal, Lacharit{\'{e}}, Brown, Li, Agarwal, and Ristenpart]{GrubbsKLBL0R20}
Paul Grubbs, Anurag Khandelwal, Marie{-}Sarah Lacharit{\'{e}}, Lloyd Brown, Lucy Li, Rachit Agarwal, and Thomas Ristenpart.
\newblock Pancake: Frequency smoothing for encrypted data stores.
\newblock In \emph{{USENIX}}, 2020.

\bibitem[Hoang et~al.(2019)Hoang, Ozmen, Jang, and Yavuz]{hoang2019hardware_posup}
Thang Hoang, Muslum~Ozgur Ozmen, Yeongjin Jang, and Attila~A Yavuz.
\newblock Hardware-supported {ORAM} in effect: Practical oblivious search and update on very large dataset.
\newblock \emph{PETS}, 2019\penalty0 (1), 2019.

\bibitem[Hoang et~al.(2020)Hoang, Behnia, Jang, and Yavuz]{hoang2020mose}
Thang Hoang, Rouzbeh Behnia, Yeongjin Jang, and Attila~A Yavuz.
\newblock Mose: Practical multi-user oblivious storage via secure enclaves.
\newblock In \emph{CODASPY}, 2020.

\bibitem[Hopcroft and Karp(1973)]{HopcroftK73}
John~E. Hopcroft and Richard~M. Karp.
\newblock An n\({}^{\mbox{5/2}}\) algorithm for maximum matchings in bipartite graphs.
\newblock \emph{SICOMP}, 2\penalty0 (4):\penalty0 225--231, 1973.

\bibitem[Intel({\natexlab{a}})]{IntelTDX}
Intel.
\newblock {MKTME} side channel impact on {Intel TDX}, {\natexlab{a}}.
\newblock URL \url{https://www.intel.com/content/www/us/en/developer/articles/technical/software-security-guidance/best-practices/mktme-side-channel-impact-on-intel-tdx.html}.

\bibitem[Intel({\natexlab{b}})]{tbb}
Intel.
\newblock Intel {oneAPI} threading building blocks ({oneTBB}), {\natexlab{b}}.
\newblock URL \url{https://github.com/oneapi-src/oneTBB}.

\bibitem[Islam et~al.(2012)Islam, Kuzu, and Kantarcioglu]{Islam2012AccessPD}
Mohammad~Saiful Islam, Mehmet Kuzu, and Murat Kantarcioglu.
\newblock Access pattern disclosure on searchable encryption: Ramification, attack and mitigation.
\newblock In \emph{{NDSS}}, 2012.

\bibitem[Ittai et~al.()Ittai, Shay, Simon, and Vincent]{Attestation}
Anati Ittai, Gueron Shay, Johnson Simon, and Scarlata Vincent.
\newblock Innovative technology for cpu based attestation and sealing.
\newblock Online at \url{https://www.intel.com/content/www/us/en/developer/articles/technical/innovative-technology-for-cpu-based-attestation-and-sealing.html}.

\bibitem[Jacob et~al.(2019)Jacob, Larsen, and Nielsen]{jacob2019lower}
Riko Jacob, Kasper~Green Larsen, and Jesper~Buus Nielsen.
\newblock Lower bounds for oblivious data structures.
\newblock In \emph{SODA}, 2019.

\bibitem[Kiefer(1953)]{kiefer1953sequential}
Jack Kiefer.
\newblock Sequential minimax search for a maximum.
\newblock \emph{AMS}, 4\penalty0 (3):\penalty0 502--506, 1953.

\bibitem[Kirsch et~al.(2010)Kirsch, Mitzenmacher, and Wieder]{kirsch2010more}
Adam Kirsch, Michael Mitzenmacher, and Udi Wieder.
\newblock More robust hashing: Cuckoo hashing with a stash.
\newblock \emph{SICOMP}, 39\penalty0 (4):\penalty0 1543--1561, 2010.

\bibitem[Komargodski and Lin(2021)]{komargodski2021logarithmic}
Ilan Komargodski and Wei-Kai Lin.
\newblock A logarithmic lower bound for oblivious {RAM} (for all parameters).
\newblock In \emph{CRYPTO}, 2021.

\bibitem[Kushilevitz et~al.(2012)Kushilevitz, Lu, and Ostrovsky]{kushilevitz2012security}
Eyal Kushilevitz, Steve Lu, and Rafail Ostrovsky.
\newblock On the (in) security of hash-based oblivious {RAM} and a new balancing scheme.
\newblock In \emph{SODA}, 2012.

\bibitem[Larsen and Nielsen(2018)]{larsen2018yes}
Kasper~Green Larsen and Jesper~Buus Nielsen.
\newblock Yes, there is an oblivious {RAM} lower bound!
\newblock In \emph{CRYPTO}, 2018.

\bibitem[Larsen et~al.(2020)Larsen, Malkin, Weinstein, and Yeo]{larsen2020lower}
Kasper~Green Larsen, Tal Malkin, Omri Weinstein, and Kevin Yeo.
\newblock Lower bounds for oblivious near-neighbor search.
\newblock In \emph{SODA}, 2020.

\bibitem[Lee et~al.(2020)Lee, Jung, Fang, Tsai, and Popa]{LeeJFTP20}
Dayeol Lee, Dongha Jung, Ian~T. Fang, Chia{-}che Tsai, and Raluca~Ada Popa.
\newblock An off-chip attack on hardware enclaves via the memory bus.
\newblock In \emph{{USENIX} Security}, 2020.

\bibitem[Lee et~al.(2017)Lee, Shih, Gera, Kim, Kim, and Peinado]{0001SGKKP17}
Sangho Lee, Ming{-}Wei Shih, Prasun Gera, Taesoo Kim, Hyesoon Kim, and Marcus Peinado.
\newblock Inferring fine-grained control flow inside {SGX} enclaves with branch shadowing.
\newblock In \emph{{USENIX} Security}, 2017.

\bibitem[Li et~al.(2014)Li, Andersen, Kaminsky, and Freedman]{li2014algorithmic}
Xiaozhou Li, David~G Andersen, Michael Kaminsky, and Michael~J Freedman.
\newblock Algorithmic improvements for fast concurrent {Cuckoo} hashing.
\newblock In \emph{EuroSys}, 2014.

\bibitem[Lin et~al.(2019)Lin, Shi, and Xie]{lin2019canosort}
Wei-Kai Lin, Elaine Shi, and Tiancheng Xie.
\newblock Can we overcome the n log n barrier for oblivious sorting?
\newblock In \emph{SODA}, 2019.

\bibitem[Lipp et~al.(2021)Lipp, Kogler, Oswald, Schwarz, Easdon, Canella, and Gruss]{LippKOSECG21}
Moritz Lipp, Andreas Kogler, David~F. Oswald, Michael Schwarz, Catherine Easdon, Claudio Canella, and Daniel Gruss.
\newblock {PLATYPUS:} software-based power side-channel attacks on x86.
\newblock In \emph{{S\&P}}, 2021.

\bibitem[Liu et~al.(2014)Liu, Zhu, Wang, and an~Tan]{Liu2014SearchPL}
Chang Liu, Liehuang Zhu, Mingzhong Wang, and Yu~an~Tan.
\newblock Search pattern leakage in searchable encryption: Attacks and new construction.
\newblock \emph{Information Sciences}, 265:\penalty0 176--188, 2014.

\bibitem[Liu et~al.(2015{\natexlab{a}})Liu, Harris, Maas, Hicks, Tiwari, and Shi]{LiuHMHTS15}
Chang Liu, Austin Harris, Martin Maas, Michael~W. Hicks, Mohit Tiwari, and Elaine Shi.
\newblock Ghostrider: {A} hardware-software system for memory trace oblivious computation.
\newblock In \emph{{ASPLOS}}, 2015{\natexlab{a}}.

\bibitem[Liu et~al.(2015{\natexlab{b}})Liu, Wang, Nayak, Huang, and Shi]{liu2015oblivm}
Chang Liu, Xiao~Shaun Wang, Kartik Nayak, Yan Huang, and Elaine Shi.
\newblock Oblivm: A programming framework for secure computation.
\newblock In \emph{S\&P}, 2015{\natexlab{b}}.

\bibitem[Lorch et~al.(2013)Lorch, Parno, Mickens, Raykova, and Schiffman]{lorch2013shroud}
Jacob~R Lorch, Bryan Parno, James Mickens, Mariana Raykova, and Joshua Schiffman.
\newblock Shroud: Ensuring private access to large-scale data in the data center.
\newblock In \emph{USENIX FAST}, 2013.

\bibitem[Lu and Ostrovsky(2013)]{lu2013distributed}
Steve Lu and Rafail Ostrovsky.
\newblock Distributed oblivious {RAM} for secure two-party computation.
\newblock In \emph{TCC}, 2013.

\bibitem[Maas et~al.(2013)Maas, Love, Stefanov, Tiwari, Shi, Asanovic, Kubiatowicz, and Song]{maas2013phantom}
Martin Maas, Eric Love, Emil Stefanov, Mohit Tiwari, Elaine Shi, Krste Asanovic, John Kubiatowicz, and Dawn Song.
\newblock Phantom: Practical oblivious computation in a secure processor.
\newblock In \emph{CCS}, 2013.

\bibitem[Maiyya et~al.(2023)Maiyya, Vemula, Agrawal, Abbadi, and Kerschbaum]{MaiyyaVAAK23}
Sujaya Maiyya, Sharath~Chandra Vemula, Divyakant Agrawal, Amr~El Abbadi, and Florian Kerschbaum.
\newblock Waffle: An online oblivious datastore for protecting data access patterns.
\newblock \emph{PACMMOD}, 1\penalty0 (4):\penalty0 266:1--266:25, 2023.

\bibitem[Matetic et~al.(2017)Matetic, Ahmed, Kostiainen, Dhar, Sommer, Gervais, Juels, and Capkun]{matetic2017rote}
Sinisa Matetic, Mansoor Ahmed, Kari Kostiainen, Aritra Dhar, David~M. Sommer, Arthur Gervais, Ari Juels, and Srdjan Capkun.
\newblock Rote: Rollback protection for trusted execution.
\newblock In \emph{USENIX Security}, 2017.

\bibitem[Mechalas and Odom()]{Mechalas2016IntelSGX}
John~P Mechalas and Benjamin~J Odom.
\newblock Intel® {Software} {Guard} extensions tutorial series: Part 1, {Intel} {SGX}.
\newblock Online at \url{https://www.intel.com/content/www/us/en/developer/articles/training/intel-software-guard-extensions-tutorial-part-1-foundation.html}.

\bibitem[Microsoft()]{AzureTEE}
Microsoft.
\newblock Azure confidential computing.
\newblock URL \url{https://learn.microsoft.com/en-us/azure/confidential-computing/}.

\bibitem[Mishra et~al.(2018)Mishra, Poddar, Chen, Chiesa, and Popa]{mishra2018oblix}
Pratyush Mishra, Rishabh Poddar, Jerry Chen, Alessandro Chiesa, and Raluca~Ada Popa.
\newblock Oblix: An efficient oblivious search index.
\newblock In \emph{S\&P}, 2018.

\bibitem[Murdock et~al.(2020)Murdock, Oswald, Garcia, Bulck, Gruss, and Piessens]{MurdockOGBGP20}
Kit Murdock, David~F. Oswald, Flavio~D. Garcia, Jo~Van Bulck, Daniel Gruss, and Frank Piessens.
\newblock Plundervolt: Software-based fault injection attacks against {Intel} {SGX}.
\newblock In \emph{S\&P}, 2020.

\bibitem[Nagarajan et~al.(2019)Nagarajan, Shafiee, Balasubramonian, and Tiwari]{NagarajanSBT19}
Chandrasekhar Nagarajan, Ali Shafiee, Rajeev Balasubramonian, and Mohit Tiwari.
\newblock {\(\rho\)}: Relaxed hierarchical {ORAM}.
\newblock In \emph{{ASPLOS}}, 2019.

\bibitem[Noble(2021)]{noble2021explicit}
Daniel Noble.
\newblock Explicit, closed-form, general bounds for {Cuckoo} hashing with a stash.
\newblock \emph{Cryptology ePrint Archive}, 2021.

\bibitem[Noble et~al.(2024)Noble, Hemenway, and Ostrovsky]{NobleHO24}
Daniel Noble, Brett Hemenway, and Rafail Ostrovsky.
\newblock {MetaDORAM}: Breaking the log-overhead information theoretic barrier.
\newblock In \emph{ECCC}, 2024.

\bibitem[Ostrovsky and Shoup(1997)]{OstrovskyS97}
Rafail Ostrovsky and Victor Shoup.
\newblock Private information storage (extended abstract).
\newblock In \emph{{STOC}}, 1997.

\bibitem[Pagh and Rodler(2004)]{pagh2004cuckoo}
Rasmus Pagh and Flemming~Friche Rodler.
\newblock Cuckoo hashing.
\newblock \emph{Journal of Algorithms}, 51\penalty0 (2):\penalty0 122--144, 2004.

\bibitem[Patel et~al.(2018)Patel, Persiano, Raykova, and Yeo]{PatelP0Y18PanORAMa}
Sarvar Patel, Giuseppe Persiano, Mariana Raykova, and Kevin Yeo.
\newblock {PanORAMa}: Oblivious {RAM} with logarithmic overhead.
\newblock In \emph{FOCS}, 2018.

\bibitem[Pinkas and Reinman(2010)]{pinkas2010oblivious}
Benny Pinkas and Tzachy Reinman.
\newblock Oblivious {RAM} revisited.
\newblock In \emph{CRYPTO}, 2010.

\bibitem[Qin et~al.(2022)Qin, Jayaram, Shi, Song, Zhuo, and Chu]{Qin22Adore}
Lianke Qin, Rajesh Jayaram, Elaine Shi, Zhao Song, Danyang Zhuo, and Shumo Chu.
\newblock Adore: Differentially oblivious relational database operators.
\newblock In \emph{VLDB}, 2022.

\bibitem[Qiu et~al.(2023)Qiu, Kellaris, Mamoulis, Nissim, and Kollios]{QiuKMNK23}
Lina Qiu, Georgios Kellaris, Nikos Mamoulis, Kobbi Nissim, and George Kollios.
\newblock Doquet: Differentially oblivious range and join queries with private data structures.
\newblock In \emph{{VLDB}}, 2023.

\bibitem[Rachid et~al.(2020)Rachid, Riley, and Malluhi]{RachidRM20page3}
Maan~Haj Rachid, Ryan~D. Riley, and Qutaibah~M. Malluhi.
\newblock Enclave-based oblivious {RAM} using {Intel}'s {SGX}.
\newblock \emph{Computers \& Security}, 91:\penalty0 101711, 2020.

\bibitem[Rane et~al.(2015)Rane, Lin, and Tiwari]{Rane2015RaccoonCD}
Ashay Rane, Calvin Lin, and Mohit Tiwari.
\newblock Raccoon: Closing digital side-channels through obfuscated execution.
\newblock In \emph{USENIX Security}, 2015.

\bibitem[Ren et~al.(2013)Ren, Yu, Fletcher, van Dijk, and Devadas]{RenYFDD13}
Ling Ren, Xiangyao Yu, Christopher~W. Fletcher, Marten van Dijk, and Srinivas Devadas.
\newblock Design space exploration and optimization of path oblivious {RAM} in secure processors.
\newblock In \emph{{ISCA}}, 2013.

\bibitem[Sasy et~al.(2018)Sasy, Gorbunov, and Fletcher]{sasy2017zerotrace}
Sajin Sasy, Sergey Gorbunov, and Christopher~W Fletcher.
\newblock {ZeroTrace}: Oblivious memory primitives from {Intel} {SGX}.
\newblock In \emph{NDSS}, 2018.

\bibitem[Sasy et~al.(2022)Sasy, Johnson, and Goldberg]{sasy2022fastocompact}
Sajin Sasy, Aaron Johnson, and Ian Goldberg.
\newblock Fast fully oblivious compaction and shuffling.
\newblock In \emph{CCS}, 2022.

\bibitem[Sasy et~al.(2023)Sasy, Johnson, and Goldberg]{Sasy0G23Waks}
Sajin Sasy, Aaron Johnson, and Ian Goldberg.
\newblock {Waks-On/Waks-Off}: Fast oblivious offline/online shuffling and sorting with {Waksman} networks.
\newblock In \emph{CCS}, 2023.

\bibitem[Seo et~al.(2017)Seo, Lee, Kim, Shih, Shin, Han, and Kim]{Seo2017SGXShieldEA}
Jaebaek Seo, Byoungyoung Lee, Seongmin Kim, Ming-Wei Shih, Insik Shin, Dongsu Han, and Taesoo Kim.
\newblock {SGX}-shield: Enabling address space layout randomization for {SGX} programs.
\newblock In \emph{NDSS}, 2017.

\bibitem[Shi(2020)]{shi2020path}
Elaine Shi.
\newblock Path oblivious heap: Optimal and practical oblivious priority queue.
\newblock In \emph{S\&P}, 2020.

\bibitem[Shih et~al.(2017)Shih, Lee, Kim, and Peinado]{shih2017t}
Ming-Wei Shih, Sangho Lee, Taesoo Kim, and Marcus Peinado.
\newblock {T-SGX}: Eradicating controlled-channel attacks against enclave programs.
\newblock In \emph{NDSS}, 2017.

\bibitem[Shinde et~al.(2016)Shinde, Chua, Narayanan, and Saxena]{ShindeCNS16page1}
Shweta Shinde, Zheng~Leong Chua, Viswesh Narayanan, and Prateek Saxena.
\newblock Preventing page faults from telling your secrets.
\newblock In \emph{AsiaCCS}, 2016.

\bibitem[Sinha et~al.(2017)Sinha, Rajamani, and Seshia]{sinha2017compiler}
Rohit Sinha, Sriram Rajamani, and Sanjit~A Seshia.
\newblock A compiler and verifier for page access oblivious computation.
\newblock In \emph{FSE}, 2017.

\bibitem[Son et~al.(2021)Son, Prechter, Poddar, Popa, and Sen]{son2021oblicheck}
Jeongseok Son, Griffin Prechter, Rishabh Poddar, Raluca~Ada Popa, and Koushik Sen.
\newblock {ObliCheck}: Efficient verification of oblivious algorithms with unobservable state.
\newblock In \emph{USENIX Security}, 2021.

\bibitem[Stefanov and Shi(2013)]{StefanovS13}
Emil Stefanov and Elaine Shi.
\newblock Oblivistore: High performance oblivious distributed cloud data store.
\newblock In \emph{{NDSS}}, 2013.

\bibitem[Stefanov et~al.(2012)Stefanov, Shi, and Song]{stefanov2011towards}
Emil Stefanov, Elaine Shi, and Dawn Song.
\newblock Towards practical oblivious {RAM}.
\newblock In \emph{NDSS}, 2012.

\bibitem[Stefanov et~al.(2018)Stefanov, Dijk, Shi, Chan, Fletcher, Ren, Yu, and Devadas]{stefanov2018path}
Emil Stefanov, Marten~van Dijk, Elaine Shi, T-H~Hubert Chan, Christopher Fletcher, Ling Ren, Xiangyao Yu, and Srinivas Devadas.
\newblock {Path ORAM}: An extremely simple oblivious {RAM} protocol.
\newblock \emph{JACM}, 65\penalty0 (4):\penalty0 1--26, 2018.

\bibitem[Tinoco et~al.(2023)Tinoco, Gao, and Shi]{tinoco2023enigmap}
Afonso Tinoco, Sixiang Gao, and Elaine Shi.
\newblock {EnigMap}: External-memory oblivious map for secure enclaves.
\newblock In \emph{USENIX Security}, 2023.

\bibitem[Tople and Saxena(2017)]{TopleS17page2}
Shruti Tople and Prateek Saxena.
\newblock On the trade-offs in oblivious execution techniques.
\newblock In \emph{{DIMVA}}, 2017.

\bibitem[Wang et~al.(2015)Wang, Chan, and Shi]{wang2015circuit}
Xiao Wang, Hubert Chan, and Elaine Shi.
\newblock Circuit {ORAM}: On tightness of the {Goldreich}-{Ostrovsky} lower bound.
\newblock In \emph{CCS}, 2015.

\bibitem[Wang et~al.(2014)Wang, Huang, Chan, Shelat, and Shi]{wang2014scoram_linear}
Xiao~Shaun Wang, Yan Huang, TH~Hubert Chan, Abhi Shelat, and Elaine Shi.
\newblock {SCORAM}: Oblivious {RAM} for secure computation.
\newblock In \emph{CCS}, 2014.

\bibitem[Weiss and Wichs(2021)]{weiss2021there}
Mor Weiss and Daniel Wichs.
\newblock Is there an oblivious {RAM} lower bound for online reads?
\newblock \emph{Journal of Cryptology}, 34\penalty0 (3):\penalty0 18, 2021.

\bibitem[Wichelmann et~al.(2024)Wichelmann, Rabich, P{\"{a}}tschke, and Eisenbarth]{WichelmannRPE24}
Jan Wichelmann, Anja Rabich, Anna P{\"{a}}tschke, and Thomas Eisenbarth.
\newblock Obelix: Mitigating side-channels through dynamic obfuscation.
\newblock In \emph{S\&P}, 2024.

\bibitem[Williams et~al.(2012)Williams, Sion, and Tomescu]{williams2012privatefs}
Peter Williams, Radu Sion, and Alin Tomescu.
\newblock Privatefs: A parallel oblivious file system.
\newblock In \emph{CCS}, 2012.

\bibitem[Wu et~al.(2023)Wu, Ning, Huang, and Liu]{WuNHL23}
Pengfei Wu, Jianting Ning, Xinyi Huang, and Joseph~K. Liu.
\newblock Differentially oblivious two-party pattern matching with sublinear round complexity.
\newblock \emph{TDSC}, 20\penalty0 (5):\penalty0 4101--4117, 2023.

\bibitem[Xu et~al.(2023)Xu, Zheng, Xu, Yuan, and Wang]{xu2023leakage}
Lei Xu, Leqian Zheng, Chengzhi Xu, Xingliang Yuan, and Cong Wang.
\newblock Leakage-abuse attacks against forward and backward private searchable symmetric encryption.
\newblock In \emph{CCS}, 2023.

\bibitem[Xu et~al.(2015)Xu, Cui, and Peinado]{XuCP15}
Yuanzhong Xu, Weidong Cui, and Marcus Peinado.
\newblock Controlled-channel attacks: Deterministic side channels for untrusted operating systems.
\newblock In \emph{S\&P}, 2015.

\bibitem[Yeo(2023)]{yeo2023cuckoo}
Kevin Yeo.
\newblock Cuckoo hashing in cryptography: Optimal parameters, robustness and applications.
\newblock In \emph{CRYPTO}, 2023.

\bibitem[Zahur et~al.(2016)Zahur, Wang, Raykova, Gasc{\'o}n, Doerner, Evans, and Katz]{zahur2016revisiting}
Samee Zahur, Xiao Wang, Mariana Raykova, Adri{\`a} Gasc{\'o}n, Jack Doerner, David Evans, and Jonathan Katz.
\newblock Revisiting square-root {ORAM}: Efficient random access in multi-party computation.
\newblock In \emph{S\&P}, 2016.

\bibitem[Zhang et~al.(2020)Zhang, Song, Yin, Zou, Shi, and Jin]{zhang2020klotski}
Pan Zhang, Chengyu Song, Heng Yin, Deqing Zou, Elaine Shi, and Hai Jin.
\newblock Klotski: Efficient obfuscated execution against controlled-channel attacks.
\newblock In \emph{ASPLOS}, 2020.

\bibitem[Zheng et~al.(2024)Zheng, Xu, Wang, Wang, Hu, Qin, Li, and Ren]{LeqianSWAT}
Leqian Zheng, Lei Xu, Cong Wang, Sheng Wang, Yuke Hu, Zhan Qin, Feifei Li, and Kui Ren.
\newblock {SWAT:} {A} system-wide approach to tunable leakage mitigation in encrypted data stores.
\newblock In \emph{VLDB}, 2024.

\bibitem[Zheng et~al.(2017)Zheng, Dave, Beekman, Popa, Gonzalez, and Stoica]{zheng2017opaque}
Wenting Zheng, Ankur Dave, Jethro~G Beekman, Raluca~Ada Popa, Joseph~E Gonzalez, and Ion Stoica.
\newblock Opaque: An oblivious and encrypted distributed analytics platform.
\newblock In \emph{NSDI}, 2017.

\bibitem[Zhuang et~al.(2004)Zhuang, Zhang, and Pande]{zhuang2004hide}
Xiaotong Zhuang, Tao Zhang, and Santosh Pande.
\newblock Hide: An infrastructure for efficiently protecting information leakage on the address bus.
\newblock \emph{ACM SIGOPS Operating Systems Review}, 38\penalty0 (5), 2004.

\end{thebibliography}
}
\appendix
\section{Preliminaries (Cont.)}\label{sec:append:oram}
All functionalities that we formalize here describe only the input-output behavior of the primitives rather than their implementation details.
The ideal functionality of an ORAM, as shown in~\cref{func:oram}, implements logical memory. 

\begin{functionality}\label{func:oram}
\caption{Oblivious RAM \func{ORAM}:}
\func{ORAM} reactively holds $n$ $w$-bit memory words $A[1,\dots, n]$, in which $A[\mathtt{addr}]$ is initialized as $0^w, \forall \mathtt{addr}\in[n]$. 
\\
$\bullet~\func{ORAM}.\mathsf{access}(\mathtt{op}, \mathtt{addr}, v)$: \Comment{{\small$\mathtt{op}\in\{\mathtt{read}, \mathtt{write}\},$ $\mathtt{addr}\in[n],$ $v\in\{0, 1\}^w$}}

\begin{algorithmic}[1]
    \If{$\mathtt{op}=\mathtt{read}$} $\mathtt{res}\gets A[\mathtt{addr}]$
    \Else~$A[\mathtt{addr}]\gets v, \mathtt{res}\gets v$    \Comment{{\small$\mathtt{op}=\mathtt{write}$}}
    \EndIf
    \State \Return $\mathtt{res}$
\end{algorithmic}
\end{functionality}

The ideal functionality shown in~\cref{func:oht} implements a hash table (also known as a dictionary) that reactively supports table build, key lookup, and data extraction operations.

\begin{functionality}\label{func:oht}
    \caption{Oblivious Hash Table \func{HT}:}
    Denote a key pair as $(k, v)\in\{0, 1\}^{\ell_k}\times\{0, 1\}^{\ell_v}$. 
    W.l.o.g., we assume that $\ell_k=\bigO{w}$ and $\ell_v=\bigO{w}$, \ie, both the key and the value can be stored in $\bigO{w}$ memory words.
    \\
    $\bullet~\func{HT}.\mathsf{build}(A)$: \Comment{{\small$A$ contains $n$ possibly dummy (\ie, $\bot$) key-value pairs with distinct keys}}
    \begin{algorithmic}[1]
        \State initialize its internal state $(A, T)$ with $T\gets \emptyset$
        \State output nothing
    \end{algorithmic}

    $\bullet~\func{HT}.\mathsf{lookup}(k)$: \Comment{{\small $k\in\{0, 1\}^{\ell_k}\cup\{\bot\}$}}
    \begin{algorithmic}[1]
        \If{$k\neq\bot \wedge k\in T$} \Return \textsf{fail} \Comment{{\small $k$ is a recurrent lookup}}
        \EndIf
        \If{$k=\bot \vee k\notin A$} $v'\gets\bot$
        \Else 
        \State $v'\gets v$, where $(k, v)\in A$
        \State $T\gets T\cup\{(k, v')\}$
        \EndIf
        \State \Return $v'$
    \end{algorithmic}

    $\bullet~\func{HT}.\mathsf{extract}()$: 
    \begin{algorithmic}[1]
        \For{$(k, v)\in A$}
        \If{$k\in T$} $k\gets \bot, v\gets \bot$
        \EndIf
        \EndFor
        \State shuffle $A$ uniformly at random
        \State \Return $A$
    \end{algorithmic}
\end{functionality}

We then formally define oblivious simulation of a (reactive) functionality \func{F} in~\cref{def:oblivious-simulation}.
W.l.o.g., we assume that \func{F} takes commands and input data of the form $(\mathtt{cmd}, \mathtt{inp})$, and produces an output $\mathtt{out}$, while probably maintaining some internal (secret) state.
For a RAM machine $M_{\mathsf{F}}$ implementing \func{F}, the execution of $(\mathtt{cmd}, \mathtt{inp})$ will produce side-channel information $\mathtt{addrs}$ that indicates the memory addresses accessed during the process. 
We occasionally refer to certain parameters as ``public'', indicating that the simulator additionally takes these parameters as input and that it is acceptable for adversaries to have knowledge of them. 
\begin{definition}[Oblivious Machine Implementing a Functionality \func{F}]\label{def:oblivious-simulation}
A RAM machine $M_{\mathsf{F}}$ \emph{obliviously implements} the reactive functionality \func{F} if for any probabilistic polynomial-time (PPT) adversary $\adv$, there exists a PPT simulator $\mathsf{Sim}$, such that the view of the adversary $\adv$ in the following experiments $\mathsf{Expt}_{\adv}^{{\normalfont real}, M_{\mathsf{F}}}(1^{\secpar})$ and $\mathsf{Expt}_{\adv, \mathsf{Sim}}^{{\normalfont ideal}, \func{F}}(1^{\secpar})$ is computationally indistinguishable.
\end{definition}
\vspace{-1em}
\begin{mdframed}[hidealllines=true, leftmargin=0pt, rightmargin=0pt, innerleftmargin=0pt, innerrightmargin=0pt]
\noindent\underline{$\mathsf{Expt}_{\adv}^{\text{real}, M_{\mathsf{F}}}(1^{\secpar})$}:
\begin{algorithmic}[1]
\State $(\mathtt{cmd}_1, \mathtt{inp}_1)\gets \adv(1^{\secpar}),~i\gets 1$
\While{$\mathtt{cmd}_i\neq \bot$}
\State \textcolor{teal}{$(\mathtt{out}_i, \mathtt{addrs}_i)\gets M_{\mathsf{F}}(1^{\secpar},\mathtt{cmd}_i, \mathtt{inp}_i)$}
\State $(\mathtt{cmd}_{i+1}, \mathtt{inp}_{i+1})\gets \adv(1^{\secpar}, \mathtt{out}_i, \mathtt{addrs}_i)$
\State $i\gets i+1$
\EndWhile
\end{algorithmic}
\vspace{1em}
\noindent\underline{$\mathsf{Expt}_{\adv, \mathsf{Sim}}^{\text{ideal}, \func{F}}(1^{\secpar})$}:
\begin{algorithmic}[1]
\State $(\mathtt{cmd}_1, \mathtt{inp}_1)\gets \adv(1^{\secpar}), i\gets 1$
\While{$\mathtt{cmd}_i\neq \bot$}
\State \textcolor{teal}{$\mathtt{out}_i\gets \func{F}(\mathtt{cmd}_i, \mathtt{inp}_i),~\mathtt{addrs}_i\gets \mathsf{Sim}(\mathtt{cmd}_i, 1^{\secpar})$}
\State $(\mathtt{cmd}_{i+1}, \mathtt{inp}_{i+1})\gets \adv(1^{\secpar}, \mathtt{out}_i, \mathtt{addrs}_i)$
\State $i\gets i+1$
\EndWhile
\end{algorithmic}
\end{mdframed}

\section{Details on \system}\label{sec:append:h2o2ram}
\balance
We provide the details of \system's implementation and the proof of \cref{thm:h2o2ram} here. 
Recall that \system consists of $\bigO{n}$ levels. %
Let $L:=\lceil\log n\rceil$ and $\ell$ be the threshold representing the highest level at which a linear scan performs optimally among all the hash schemes presented in~\cref{tab:ohashes}. 
Each level $i\in\{\ell, \dots, L\}$ is a tailored hash table $T_i$ with a capacity of $2^i$, where $T_{\ell}$ is specifically implemented as a plain array. 
\begin{algorithm}
    \caption{$\system.\mathsf{access}(\mathtt{op}, \mathtt{addr}, v)$:}\label{alg:h2o2ram}
    \begin{algorithmic}[1]
        \Require  $\mathtt{op}\in\{\mathtt{read}, \mathtt{write}\},$ $\mathtt{addr}\in[n],$ $v\in\{0, 1\}^w$
        \State initialize $\mathtt{res}\gets \bot$ 
        \For {$i\in\{\ell, \dots, L\}$}
        \If{$T_i$ is empty} continue
        \EndIf 
        \If {$\mathtt{res}=\bot$}
            $\mathtt{res}\gets T_i.\mathsf{lookup}(\mathtt{addr})$
        \Else~$T_i.\mathsf{lookup}(\bot)$ %
        \EndIf
        \EndFor
        \If {$\mathtt{op}=\mathtt{write}$}
            $\mathtt{res}\gets v$
        \EndIf
        \State append $(\mathtt{addr}, \mathtt{res})$ to $T_{\ell}$ \Comment{{\small $T_{\ell}$ is a plain array}}
        \If {$T_{\ell}$ is full} 
        \State let $i^*$ be the first empty level or $i^*=L$ if all levels are non-empty
        \State let $A\gets T_{\ell}.\mathsf{extract}()||\dots||T_{i^*-1}.\mathsf{extract}()$
        \If {$i^*=L$}\label{alg:h2o2ram:extract}
        obliviously compact $A$ to its half
        \EndIf 
        \State obliviously intersperse (\ie, shuffle) $A$
        \State $T_{i^*}.\mathsf{build}(A)$
        \EndIf
        \State \Return $\mathtt{res}$
    \end{algorithmic}
\end{algorithm}

For brevity, we omit precise negligible probabilities of certain oblivious building blocks, instead denoting them as $\delta_{\mathsf{oht}}$, $\delta_{\mathsf{oshuffle}}$, $\delta_{\mathsf{ocompact}}$, and $\delta_{\mathsf{ointersperse}}$ corresponding to the oblivious hash table, oblivious shuffle, oblivious compact, and oblivious intersperse operations, respectively.
The mathematical forms of these values can be derived by examining their specific instantiations.
For instaince, the ocompact implementation~\cite{asharov2023futorama} yields $\delta_{\mathsf{ocompact}}=\bigO{\frac{n}{\mathsf{poly}{\log n}}e^{-\mathsf{poly}{\log n}}}$. %
\Cref{thm:ocuckoo} shows that our oblivious stashless cuckoo hash has $\delta_{\mathsf{oht\_cuckoo}}=n^{-\mathsf{poly}(k^2)}+e^{-\bigO{kn}}$ with $k=\omega(1)$.

Let $\mathsf{Sim}_{\mathsf{F}}$ denote the oblivious simulator of a functionality \func{F}, $n$, $L$, and $\ell$ be defined as the ones in \system. 
We build the oblivious simulator of \system as shown in~\cref{alg:h2o2ram:sim}. 
\begin{osimulator}\label{alg:h2o2ram:sim}
    \caption{$\mathsf{Sim}_{\system}(\mathsf{access}, 1^{\secpar}):$}
    Mark all levels but the bottom as empty.
\begin{algorithmic}[1]
    \Ensure memory access pattern $\mathtt{addrs}$
    \State initialize $\mathtt{addrs}\gets \emptyset$
    \For {$i\in\{\ell, \dots, L\}$}
    \If{level $i$ is empty} continue \Comment{\small public information}
    \EndIf 
    \State$\mathtt{addrs}\gets \mathtt{addrs}\cup\{\mathsf{Sim}_{\mathsf{HT}_i}(\mathsf{lookup}, 1^{\secpar})\}$
    \EndFor
    \State add the addresses of $\mathtt{op}, \mathtt{res}, v$, and the tail of $T_{\ell}$ to $\mathtt{addrs}$
    \If {level $\ell$ is full}  \Comment{\small public information}
    \State let $i^*$ be the first empty level or $i^*=L$ if all levels are non-empty \Comment{\small public information}
    \State $\mathtt{addrs} \gets \mathtt{addrs}\cup\{\mathsf{Sim}_{\mathsf{HT}_{\ell}}(\mathsf{extract}, 1^{\secpar})$ $||$ $\dots$ $||$ $\mathsf{Sim}_{\mathsf{HT}_{i^*-1}}(\mathsf{extract}, 1^{\secpar})\}$
    \State let $\tilde{n}$ be the length of $A$ in~\cref{alg:h2o2ram}~\cref{alg:h2o2ram:extract} %
    \If {$i^*=L$}
    \State $\mathtt{addrs} \gets \mathtt{addrs}\cup\{{ \mathsf{Sim}_{\mathsf{ocompact}}(1^{\secpar},\tilde{n})}\}$
    \EndIf 
    \State $\mathtt{addrs} \gets \mathtt{addrs}\cup\{{ \mathsf{Sim}_{\mathsf{ointersperse}}(1^{\secpar}, \tilde{n})}\}$ 
    \State  $\mathtt{addrs} \gets \mathtt{addrs}\cup\{{ \mathsf{Sim}_{\mathsf{HT}_{i^*}}(\mathsf{build}, 1^{\secpar}, \tilde{n})}\}$ 
    \State mark the levels from $\ell$ to $i^*$ as empty and $i^*$ as non-empty
    \EndIf
    \State \Return $\mathtt{addrs}$
\end{algorithmic}
\end{osimulator}
To prove \cref{thm:h2o2ram}, the remaining task is to construct hybrid constructions between \cref{alg:h2o2ram} (denoted as Construction $1$) and \cref{alg:h2o2ram:sim}. 
Construction $2$ is the same as Construction $1$ except that lines $4\sim5$ are replaced by the line $4$ in $\mathsf{Sim}_{\system}$, yielding an adversarial advantage of $\sum_{i\in\{\ell, \dots, L\}}\delta_{{\mathsf{HT}_i}(\mathsf{lookup})}$. 
Construction $3$ is the same as Construction $2$ except that lines $6\sim 7$ are replaced by the line $7$ in $\mathsf{Sim}_{\system}$. 
Construction $4$ is the same as Construction $3$ except that line $10$ is replaced by the line $8$ in $\mathsf{Sim}_{\system}$, yielding an adversarial advantage of $\sum_{i\in\{\ell, \dots, L\}}\delta_{{\mathsf{HT}_i}(\mathsf{extract})}$. 
Construction $5$ is the same as Construction $4$ except that lines $11\sim12$ are replaced by lines $9\sim 13$ in $\mathsf{Sim}_{\system}$, yielding an adversarial advantage of $\delta_{{\mathsf{ocompact}}} + \delta_{{\mathsf{ointersperse}}}$. 
Finally, \cref{alg:h2o2ram:sim} is the same as Construction $5$ except that the line $13$ in \cref{alg:h2o2ram} is replaced by its line $13$, yielding an adversarial advantage of $\sum_{i\in\{\ell, \dots, L\}}\delta_{{\mathsf{HT}_i}(\mathsf{build})}$. 
In short, the adversarial advantage of \system is upper-bounded by $\sum_{i\in\{\ell, \dots, L\}}\delta_{{\mathsf{HT}_i}} + \delta_{{\mathsf{ocompact}}} + \delta_{{\mathsf{ointersperse}}}$. 
As all the above values are negligible, the advantage of any PPT adversary against \system is also negligible. 
It concludes the proof of \cref{thm:h2o2ram}. 

\end{document}